\documentclass[aps,prc,twocolumn,superscriptaddress]{revtex4-1}

\usepackage{graphicx}
\usepackage{amsmath}
\usepackage{braket}
\usepackage{url}
\usepackage{multirow}
\usepackage{multirow}
\usepackage{amssymb}
\usepackage{booktabs}
\usepackage{rotating}
\usepackage{bm}
\usepackage{bbm}
\usepackage{ulem}

\usepackage[colorlinks,
linkcolor=blue,
anchorcolor=blue,
urlcolor=blue,
citecolor=blue]{hyperref}

\usepackage[mathlines]{lineno}

\begin{document}


\newcommand{\rwYQ}[1]{ {\color{green} #1}}

\title{Deformed in-medium similarity renormalization group}

\author{Q. Yuan}
 \affiliation{%
School of Physics, and State Key Laboratory of Nuclear Physics and Technology, Peking University, Beijing 100871, China
}%
\author{S. Q. Fan}
 \affiliation{%
School of Physics, and State Key Laboratory of Nuclear Physics and Technology, Peking University, Beijing 100871, China
}%
\author{B. S. Hu}
 \affiliation{%
School of Physics, and State Key Laboratory of Nuclear Physics and Technology, Peking University, Beijing 100871, China
}%
\author{J. G. Li}
 \affiliation{%
School of Physics, and State Key Laboratory of Nuclear Physics and Technology, Peking University, Beijing 100871, China
}%
\author{S. Zhang}
 \affiliation{%
School of Physics, and State Key Laboratory of Nuclear Physics and Technology, Peking University, Beijing 100871, China
}%
\author{S. M. Wang}
 \affiliation{%
Key Laboratory of Nuclear Physics and Ion-beam Application (MOE), Institute of Modern Physics, Fudan University, Shanghai 200433, China
}
\affiliation{%
Shanghai Research Center for Theoretical Nuclear Physics, NSFC and Fudan University, Shanghai 200438, China
}%
\author{Z. H. Sun}
 \affiliation{%
Physics Division, Oak Ridge National Laboratory, Oak Ridge, Tennessee 37831, USA
}%
\author{Y. Z. Ma}
 \affiliation{%
School of Physics, and State Key Laboratory of Nuclear Physics and Technology, Peking University, Beijing 100871, China
}%
\author{F. R. Xu}%
 \email{frxu@pku.edu.cn}
\affiliation{%
School of Physics, and State Key Laboratory of Nuclear Physics and Technology, Peking University, Beijing 100871, China
}%


\begin{abstract}
We have developed an {\it ab initio} deformed in-medium similarity renormalization group (IMSRG) for open-shell nuclei. This is a single-reference IMSRG in deformed Hartree-Fock (HF) basis. Deformed wave functions are more efficient in describing deformed nuclei. The broken spherical symmetry needs to be restored by angular momentum projection, which is computational expensive. The angular momentum mainly capture the static correlations and can be estimated by the projection of the HF state. In this work, we do deformed IMSRG calculation and add the correlation energy from projected HF as a leading order approximation. As the test ground, we have calculated the deformed $^{8,10}\rm Be$ isotopes with the optimized chiral interaction NNLO$_{\rm opt}$. The results are benchmarked with the no-core shell model and valence space IMSRG calculations. Then we systematically investigated the ground-state energies and charge radii of even-even isotopes from light beryllium to medium-mass magnesium. The calculated energies are extrapolated to infinite basis space by an exponential form, and compared with the extrapolated valence-space IMSRG results and experimental data available. 
\end{abstract}{}

\maketitle
The {\it ab initio} calculation of nuclei is the frontier of current nuclear physics theory. In the past two decades, many progresses have been made in {\it ab initio} many-body methods and internucleon interactions. The no-core shell model (NCSM) \cite{Navratil-PhysRevC.62.054311(2000),BARRETT2013131} and quantum Monte Carlo \cite{RevModPhys.87.1067} can provide exact solutions to light nuclei. {\it Ab initio} in-medium similarity renormalization group (IMSRG) \cite{IMSRG-PhysRevLett.106.222502(2011),IMSRG-Phys.Rep.621.165(2016)}, coupled cluster (CC)~\cite{CC-Rep.Prog.Phys.77.096302(2014),CC-PhysRevLett.113.142502(2014)}, self-consistent Green's function \cite{W.H.Dickhoff-Prog.Part.Nucl.Phys.52.377(2004),V.Soma-PhysRevC.87.011303(2013)} and many-body perturbation theory (MBPT)~\cite{A.Tichai-Phys.Lett.B.756.283(2016),B.S.Hu-PhysRevC.94.014303(2016),A.Tichai-Phys.Lett.B.786.195(2018)} can go to heavier mass regions. 

In the spherical $j$-scheme with a single reference state, IMSRG \cite{IMSRG-PhysRevLett.106.222502(2011),IMSRG-Phys.Rep.621.165(2016)}, CC \cite{CC-PhysRevLett.101.092502(2008), CC-Rep.Prog.Phys.77.096302(2014)} and MBPT~\cite{A.Tichai-Phys.Lett.B.756.283(2016),B.S.Hu-PhysRevC.94.014303(2016),A.Tichai-Phys.Lett.B.786.195(2018)} work only for closed-shell nuclei. For an open-shell nucleus, one cannot write the single reference state due to the degeneracy of the last single$\text{-}j$ orbits. To overcome this shortcoming in open-shell nuclei, the multi-reference versions of IMSRG~\cite{MRIMSRG-PhysRevLett.110.242501(2013),MRIMSRG-PhysRevC.90.041302(2014)}, CC~\cite{HFBCC-PhysRevC.91.064320(2015)} and MBPT~\cite{mMBPT1, mMBPT2, mMBPT3} have been developed using a spherical Hartree-Fock-Bogoliubov (HFB) quasiparticle reference state which is projected with the good numbers of protons and neutrons. This, of course, significantly complicates the formalism and increases the computational cost. Alternatively, one can derive a valence-space effective interaction using the spherical single-reference IMSRG~\cite{IMSRG-PhysRevLett.113.142501(2014),IMSRG-PhysRevLett.118.032502(2017)}, single-reference CC~\cite{CC-PhysRevLett.113.142502(2014),CC-PhysRevC.98.054320(2018)} or single-reference MBPT~\cite{M.Hjorth.Jensen-Phys.Rep.261.125(1995), L.Coraggio-Ann.Phys.327.2125(2012), Z.H.Sun-Phys.Lett.B.769.227(2017), B.S.Hu-Phys.Lett.B.802.135206(2020)} at a shell closure, and perform shell model calculations for open-shell nuclei. 

Using the $m$-scheme, however, one can write a single Hartree-Fock (HF) reference state for any even-even nuclei. This corresponds to the introduction of deformation degree of freedom. Recently, the deformed CC \cite{Duguet_2014} has been developed without~\cite{DCC-PhysRevC.102.051303(2020)} and with the angular momentum projection \cite{PCC-arXiv.2201.07298(2022)}, showing that the deformed single-reference CC is well powerful in describing open-shell nuclei~\cite{DCC-PhysRevC.102.051303(2020),DCC-PhysRevLett.126.182502(2021), PCC-arXiv.2201.07298(2022)}. The reference state with the rotational symmetry broken could better reflect the intrinsic structure of the deformed nucleus, and captures more correlations which would be many-particle$-$many-hole excitations in the spherical scheme through symmetry restoration \cite{book}. The preconsideration of expected symmetries, e.g., as done in the symmetry-adapted approach~\cite{T.Dytrych-PhysRevLett.111.252501(2013), T.Dytrych-PhysRevLett.124.042501(2020)}, can provide an efficient way to capture the expected features of nuclear states of interest and, on the other hand, can reduce the computational cost. 


As one of the powerful $ab\ initio$ methods, the IMSRG formulated in terms of continuous unitary transformation provides an efficient tool to treat energy and other observables equally. Its extension to a deformed scheme should be useful for the descriptions of deformed open-shell nuclei. In this paper, we present the single-reference deformed IMSRG (D-IMSRG) in the deformed HF basis. $^8\rm Be$ and $^{10}\rm Be$ are exotic with the structure of $2\alpha$ cluster or elongated shape. We perform the D-IMSRG calculations for $^{8,10}\rm Be$ ground-state energies, and benchmark with the NCSM and valence-space IMSRG (VS-IMSRG) results. Then, we apply the D-IMSRG to the ground-state energies and charge radii of even-even nuclei from light Be to medium-mass Mg isotopes. 


We start from an intrinsic $A$-body Hamiltonian which is normal ordered with respect to the deformed $A$-dependent reference state $|\Phi\rangle$ (i.e., the $m$-scheme HF ground state of the target nucleus). The normal ordered intrinsic Hamiltonian reads
\begin{equation}
H=E_0+\sum_{ij}f_{ij}:a^{\dagger}_ia_j:+\frac{1}{2!^2}\sum_{ijkl}\Gamma_{ijkl}:a^{\dagger}_ia^{\dagger}_ja_la_k: ,
\label{eq1}
\end{equation}
where $E_0$, $f$, and $\Gamma$ correspond to the normal ordered zero-, one-, and two-body terms, respectively. In the present work, we use the optimized chiral nucleon-nucleon (NN) interaction $\rm NNLO_{\rm opt}$ \cite{A.Ekstrom-PhysRevLett.110.192502(2013),R.Kanungo-PhysRevLett.117.102501(2016),B.S.Hu-PhysRevC.99.061302(2019)}, which gives good descriptions of nuclear binding energies, excitation spectra and neutron matter equation of state without resorting to 3NFs.

First we solve the axially deformed HF equation of the even-even nucleus within the spherical harmonic oscillator (HO) basis. The deformed HF single-particle levels obtained are twofold degenerate with respect to the angular momentum projection quantum number $m$ of the orbital (i.e., the energies are the same with respect to $\pm m$). We fill the deformed HF single-particle levels up to the Fermi surfaces of neutrons and protons in $\pm m$ pairing from low to high $|m|$, which keeps the axial, parity and time-reversal symmetries of the even-even ground state, creating a prolate deformed HF reference state~\cite{DCC-PhysRevC.102.051303(2020)}.  

The aim of the IMSRG is to drive the many-body Hamiltonian into a band- or block-diagonal form, using continuous unitary transformation. Hamiltonian~(\ref{eq1}) can be written in the diagonal $H^d(s)$ and off-diagonal $H^{od}(s)$ parts, 
\begin{equation}
{H(s)=U(s)H(0)U^{\dagger}(s)\equiv H^d(s)+H^{od}(s).}
\end{equation}
With the continuous unitary transformation $U(s)$, we aim for $\lim _{s \rightarrow \infty} H^{od}(s)=0$. In practice, the transformation is achieved by solving the flow equation,
\begin{equation}
\frac{dH(s)}{ds}=[\eta(s),H(s)],
\label{FE}
\end{equation}
with an anti-Hermitian generator,
\begin{equation}
\eta(s)\equiv\frac{dU(s)}{ds}U^{\dagger}(s)=-\eta^{\dagger}(s).
\end{equation}

In Eq.~(\ref{FE}), the flow equation is truncated at the normal-ordered two-body level, which is referred to as IMSRG(2)~\cite{IMSRG-PhysRevLett.106.222502(2011)}. In this approximation, it is a simple way to define $H^{od}$ to be composed of all one- and two-body operators that connect hole (h) and particle (p) states with the deformed HF reference state, e.g., $H^{od}=\{f_{ph}, \Gamma_{pp'hh'}\}$ plus Hermitian conjugates \cite{IMSRG-Phys.Rep.621.165(2016)}. The White generator $\eta(s)$ is adopted to suppress the off-diagonal coupling $H^{od}$ to zero with a decay scale $(s-s_0)$~\cite{White-J.Chem.Phys.117.7472(2002),IMSRG-Phys.Rep.621.165(2016)}. The transformed Hamiltonian $\tilde{H}$ and other observable $\tilde{O}$ can be constructed by the Magnus definition $U=e^{\Omega}$ \cite{PhysRevC.92.034331},
\begin{equation}
\tilde{H}=e^{\Omega}He^{-\Omega}=H+[\Omega,H]+\frac{1}{2!}[\Omega,[\Omega,H]]+\dots,
\end{equation}
\begin{equation}
\tilde{O}=e^{\Omega}Oe^{-\Omega}=O+[\Omega,O]+\frac{1}{2!}[\Omega,[\Omega,O]]+\dots.
\end{equation}
Then, the ground-state energy and other observables can be calculated with the D-IMSRG ground-state wave function $|\Psi\rangle=e^{-\Omega}|\Phi\rangle$ (here $|\Phi\rangle$ is the deformed HF reference state of the nucleus) i.e., 
\begin{equation}
 E=\langle\Psi|H|\Psi\rangle= \langle\Phi|e^{\Omega}He^{-\Omega}|\Phi\rangle=\langle\Phi|\tilde{H}|\Phi\rangle,
\end{equation}
\begin{equation}
 O=\langle\Psi|O|\Psi\rangle= \langle\Phi|e^{\Omega}Oe^{-\Omega}|\Phi\rangle=\langle\Phi|\tilde{O}|\Phi\rangle.
\end{equation}

However, the exact symmetry restoration of the D-IMSRG wave function is computationally too cumbersome and expensive due to the exponential increase of configurations in projecting the wave function $|\Psi\rangle=e^{-\Omega}|\Phi\rangle$. In the deformed CC without the angular momentum projection, it was estimated that the projection of the HF state lowers the HF energy by about 3-5 MeV in the $\it sd$ shell~\cite{DCC-PhysRevC.102.051303(2020)}, which corresponds to the static correlation and is not size extensive. The modern {\it ab initio} calculations already include some of the correlations that are associated with the projection, therefore, the projection of the {\it ab initio} wave function would lead to a slightly smaller energy correction than the HF projection correction~\cite{DCC-PhysRevC.102.051303(2020), PCC-arXiv.2201.07298(2022)}. In the present work, we consider the angular momentum projection effect by taking the HF projection correction. The projection correction to the ground-state energy is estimated by
\begin{equation}
{\Delta E_{\rm proj}= \frac{\langle\Phi|HP|\Phi\rangle}{\langle\Phi|P|\Phi\rangle}-\frac{\langle\Phi|H|\Phi\rangle}{\langle\Phi|\Phi\rangle},}
\end{equation}
where $P^J_{MM'}=\frac{2J+1}{8\pi^2}\int d\omega D^{J*}_{MM'}(\omega)R(\omega)$ is the angular momentum projection operator. This provides a D-IMSRG ground-state energy given by $E+\Delta E_{\rm proj}$ with the projection correction estimated by the HF wave function (here $E$ is obtained by Eq.~(7), i.e., the ground-state energy without the projection).


\begin{figure*}[ht]
\includegraphics[width=0.95\textwidth]{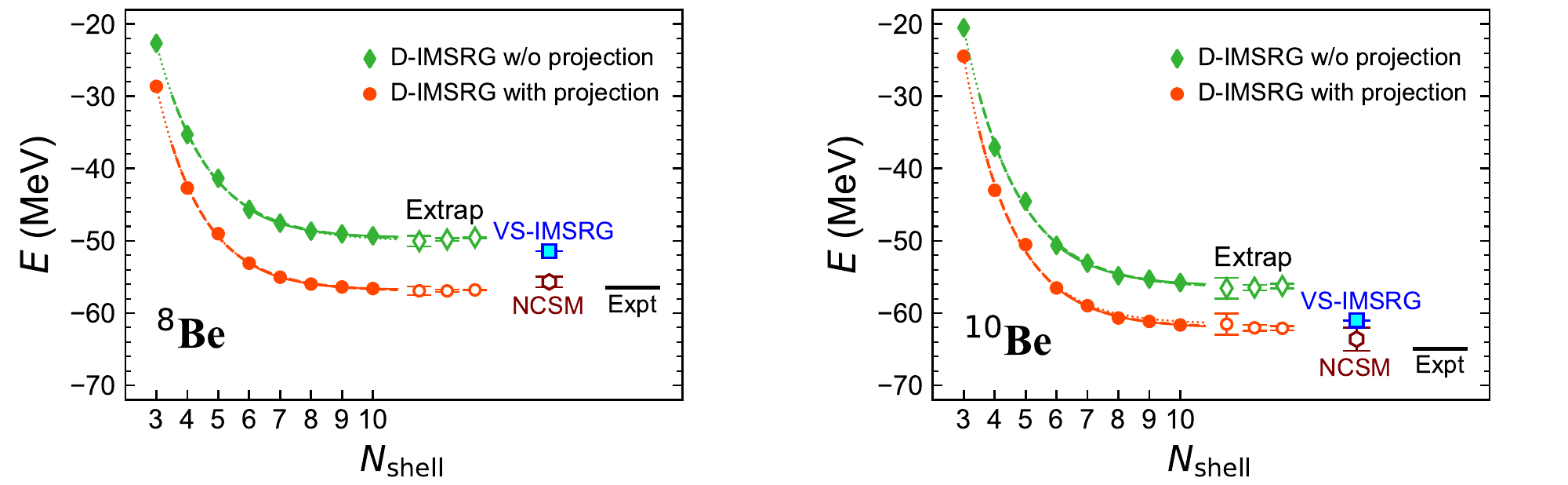}
\caption{\label{fig1} Ground-state energies calculated by D-IMSRG with and without the projection correction for $^8\rm Be$ and $^{10}\rm Be$, with respect to the basis-space size $N_{\rm shell}$. Symbols below ``Extrap" represent the exponential extrapolated energies to infinite basis space, based on different data points, from left to right, $N_{\rm shell}=$3-7, 3-10 and 6-10, respectively, with the fitting uncertainties given by error bars. Extrapolation uncertainties in NCSM and VS-IMSRG calculations are also given by error bars there. Experimental data are taken from AME2020 \cite{Wang_2021}.}
\end{figure*}


 In the spherical $j$-scheme, single-particle levels of the same $j$-shell are degenerate. The degeneracy is broken with deformation appearing, though a twofold degeneracy with respect to $\pm m$ remains at an axially symmetric shape. This increases dramatically the model-space dimension. The D-IMSRG space dimension depends on the nucleon number $A$ and the basis-space size $N_{\rm shell}$ (the number of spherical HO major shells considered in solving the deformed HF). We have checked that the number of D-IMSRG Hamiltonian matrix elements in $^{40}$Mg is already over $10^9$ at $N_{\rm shell}=10$. However, such a large model space may still not be sufficient to make the calculation converged. To estimate the converged ground-state energy, we have used a simple exponential fit with respect to $N_{\rm shell}$ to extrapolate the D-IMSRG result to an infinite basis space, as done in, e.g., NCSM-type~\cite{R.Roth-PhysRevLett.99.092501(2007),R.Roth-PhysRevC.79.064324(2009),R.Roth-PhysRevLett.107.072501(2011),M.A.Caprio-J.Mod.Phys.E.24.1541002(2015),T.Abe-PhysRevC.104.054315(2021)} and multi-reference IMSRG~\cite{MRIMSRG-PhysRevLett.110.242501(2013)} calculations,
\begin{equation}
E(N_{\rm shell})=b_0+b_1exp(-b_2N_{\rm shell}),
\label{Extra}
\end{equation}
where $b_0$, $b_1$ and $b_2$ are parameters of the fit. The value of  $b_0\equiv E(N_{\rm shell}\to\infty)$ provides the estimate of the fully converged energy.

The chiral $\rm NNLO_{\rm opt}$ interaction has been used with $\hbar\omega=24$ MeV. The D-IMSRG calculations were performed within the maximum basis-space size $N_{\rm shell}=10$ of our computational feasibility. Figure \ref{fig1} shows the calculated ground-state energies of $^8\rm Be$ and $^{10}\rm Be$ with respect to the basis-space size $N_{\rm shell}$, with and without the approximate angular momentum projection. As seen in NCSM~\cite{R.Roth-PhysRevLett.99.092501(2007),R.Roth-PhysRevC.79.064324(2009),R.Roth-PhysRevLett.107.072501(2011),M.A.Caprio-J.Mod.Phys.E.24.1541002(2015),T.Abe-PhysRevC.104.054315(2021)} and multi-reference IMSRG~\cite{MRIMSRG-PhysRevLett.110.242501(2013)} calculations, the calculated energy has an exponential form with respect to the basis-space size. In Fig.~\ref{fig1}, we give the exponential extrapolated energies to infinite model space. In the exponential fit of the extrapolation using Eq.~(\ref{Extra}), different data points have been tested with the first 5 points (i.e., $N_{\rm shell}=3{\text-}7$, indicated by dotted line in Fig.~\ref{fig1}), with all the points ($N_{\rm shell}=3{\text-}10$, dashed line) and with the last 5 points ($N_{\rm shell}=6{\text-}10$, solid line). The results are given below "Extrap" in Fig.~\ref{fig1}. As shown in the figure, the extrapolations based on different data points follow almost the same exponential form (the dotted, dashed and solid lines well overlap), and the calculations with $N_{\rm shell}=10$ are almost converged in $^8\rm Be$ and $^{10}\rm Be$. The three extrapolated energies are close to each other. These indicate that the exponential extrapolation is valid. The uncertainty for the extrapolated energy can be estimated by the mean square deviations of the fit parameters in the least square fitting procedure, shown by error bars in Fig.~\ref{fig1}. We see that the uncertainty is smaller when data points with larger $N_{\rm shell}$ are fitted. 

In Fig.~\ref{fig1}, we see that the angular momentum projection corrections are  $-7.2$ and $-5.9$ MeV for $^8\rm Be$  and  $^{10}\rm Be$, respectively. The two nuclei have elongated (or $2\alpha$ cluster) shapes, therefore the projection correction is significant. The energies with the projection correction agree well with experimental data and NCSM calculations. The NCSM and VS-IMSRG calculations shown in Fig.~\ref{fig1} take the same interaction $\rm NNLO_{\rm opt}$, and are also extrapolated to infinite model space using the exponential fit. For the VS-IMSRG calculation, the $0p_{3/2,1/2}$ model space was chosen for both valence protons and neutrons outside the $^4$He core.  We find that the VS-IMSRG result underestimates the ground-state energy in $^{8,10}\rm Be$. This may be due to the missing of higher-order collective excitations which are not well treated in VS-IMSRG at the IMSRG(2) level, as discussed in Refs.~\cite{J.Henderson-Phys.Lett.B.782.468(2018),J.Henderson-arxiv.2005.03796(2020)}.

\begin{figure}[h]
\centering
  \includegraphics[width=0.42\textwidth]{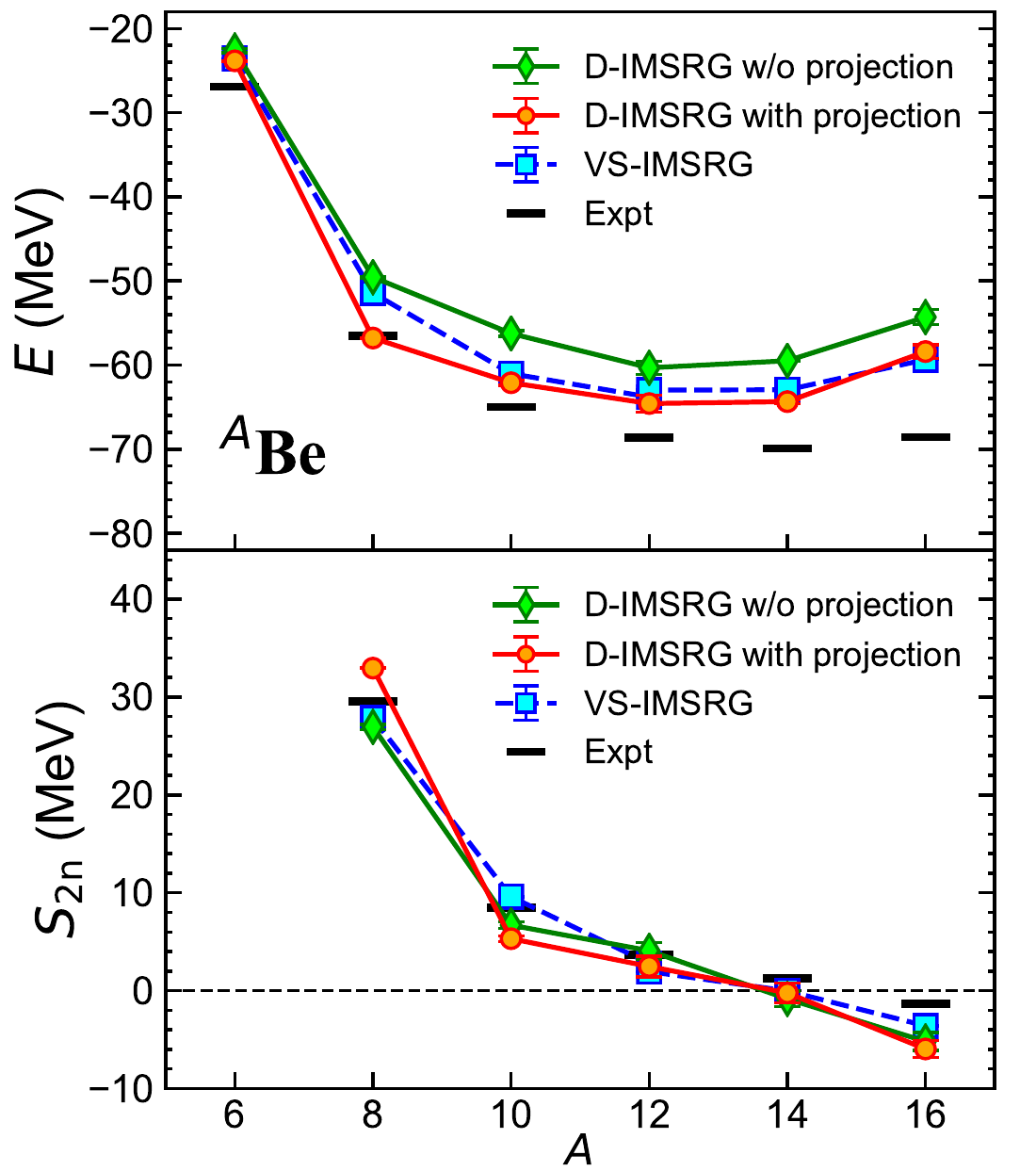}
  \caption{\label{fig2} $^{6{\text-}16}\rm Be$ ground-state energies (upper panel) and two-neutron separation energies $S_{\rm 2n}$ (lower panel) calculated by D-IMSRG with and without the projection correction, compared with VS-IMSRG calculations and experimental data \cite{Wang_2021}.}
\end{figure}

\begin{figure*}[ht]
\includegraphics[width=0.96\textwidth]{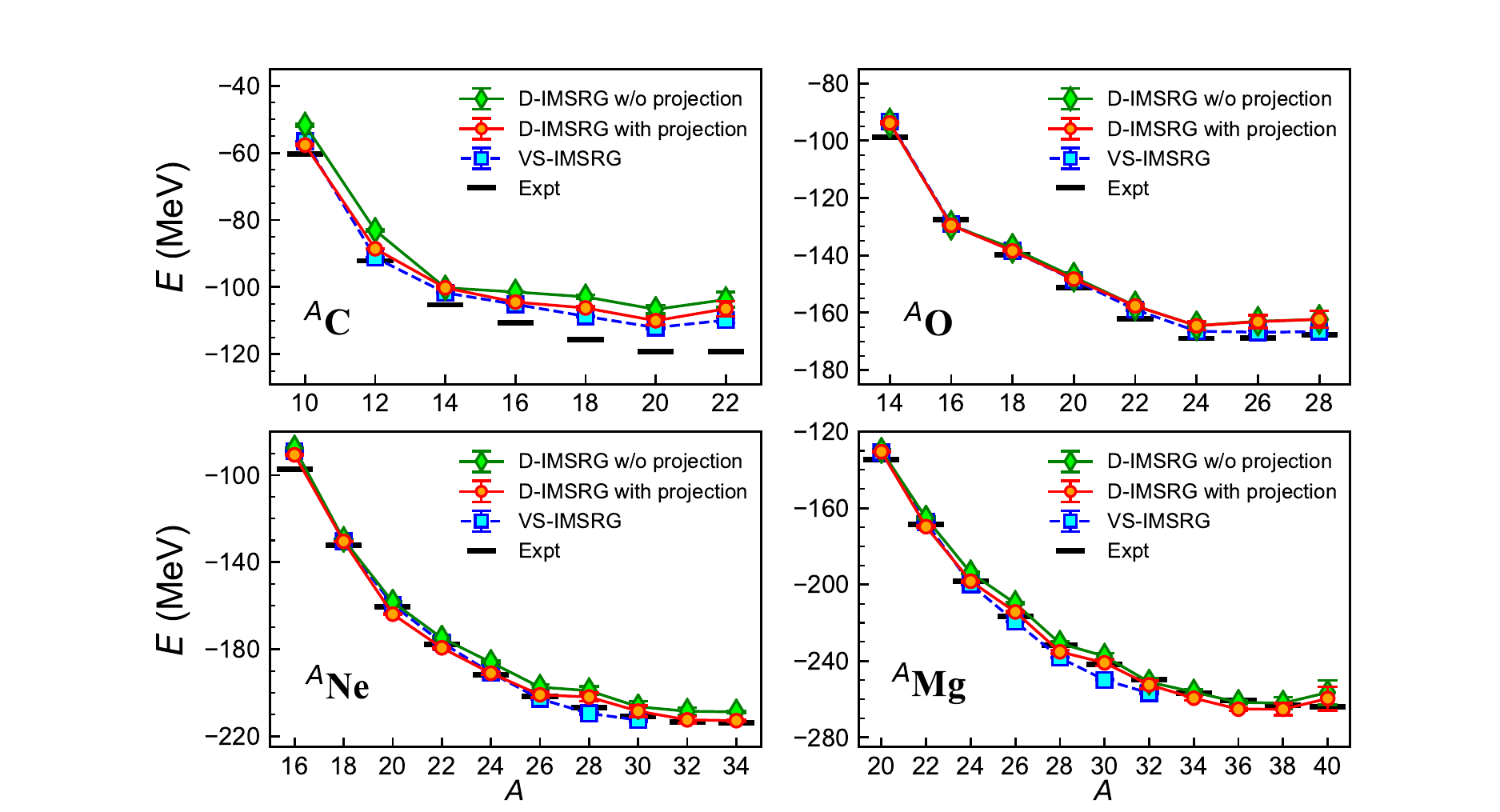}
\caption{\label{fig3} Ground-state energies of C, O, Ne and Mg isotopes. D-IMSRG results are extrapolated to infinite basis space based on $N_{\rm shell}=6{\text-}10$ data points, and the VS-IMSRG results are extrapolated based on $N_{\rm shell}=8{\text-}13$. The model space of VS-IMSRG calculations is both protons and neutrons in $0p_{3/2,1/2}$ for $^{6{\text-}14}\rm C$, protons in $0p_{3/2,1/2}$ and neutrons in $1s_{1/2}0d_{5/2,3/2}$ for $^{14{\text-}22}\rm C$, both protons and neutrons in $1s_{1/2}0d_{5/2,3/2}$ for O, Ne and Mg isotopes. Experimental data are taken from AME2020 \cite{Wang_2021}.}
\end{figure*}

\begin{figure*}[ht]
  \includegraphics[width=0.98\textwidth]{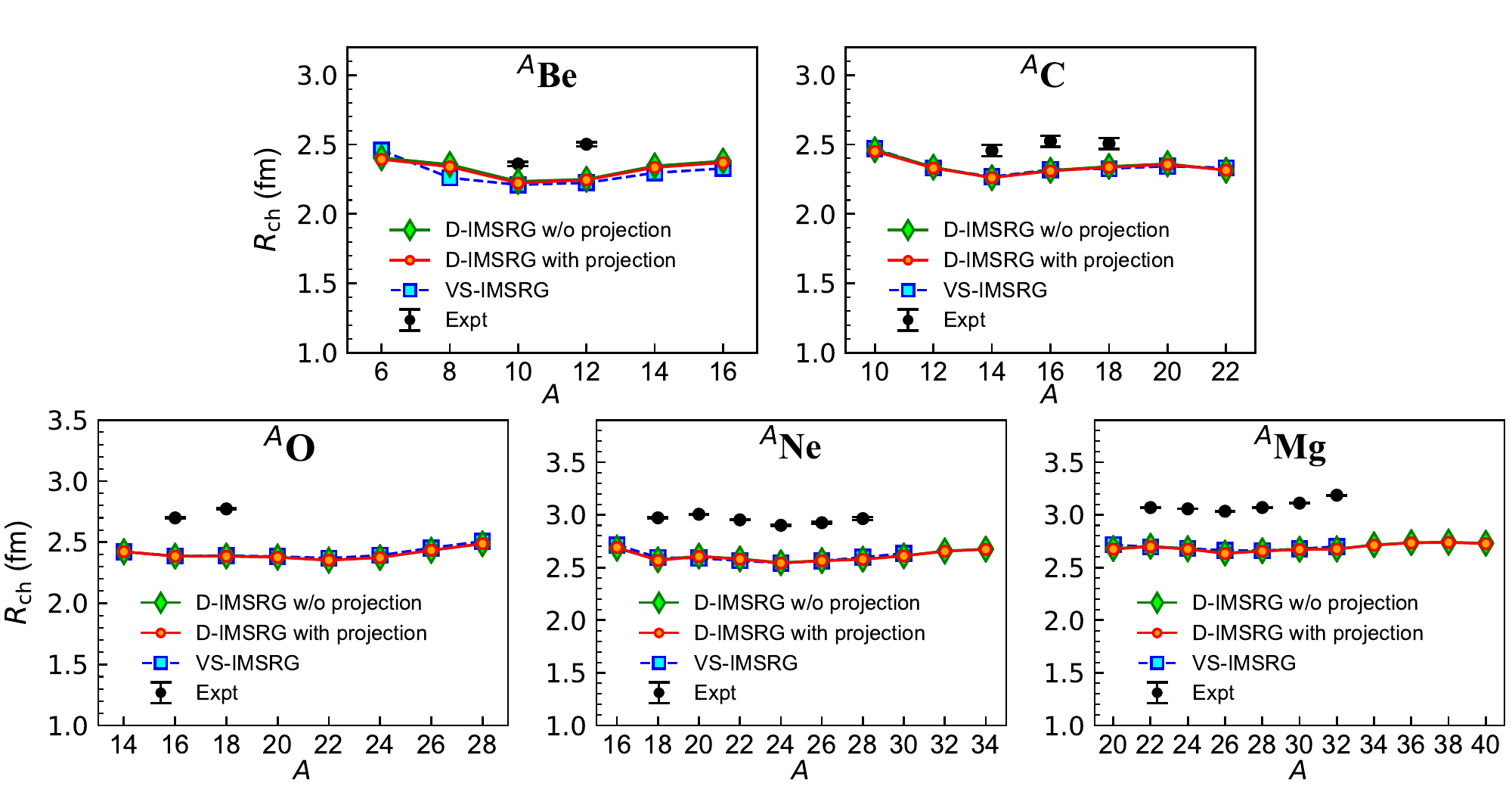}
  \caption{\label{fig4} Charge radii of Be, C, O, Ne and Mg isotopes, calculated by D-IMSRG in a basis space with $N_{\rm shell}=10$ and by VS-IMSRG with $N_{\rm shell}=13$. Experimental data are taken from \cite{A.Krieger-Appl.Phys.B.123.0(2016),R.Kanungo-PhysRevLett.117.102501(2016),V.Lapoux-PhysRevLett.117.052501(2016),B.Ohayon-PhysRevA.99.042503(2019),D.Yordanov-PhysRevLett.108.042504(2012)}.}
\end{figure*}

In Fig. \ref{fig2}, we show the calculated D-IMSRG ground-state energies (upper panel) and two-neutron separation energies (lower panel) for beryllium isotopes from  $^6\rm Be$ to $^{16}\rm Be$, along with VS-IMSRG calculations and experimental data. The D-IMSRG energies are extrapolated based on $N_{\rm shell}=6{\text -}10$ data points, while the VS-IMSRG results are extrapolated based on $N_{\rm shell}=8{\text -}13$ data points (VS-IMSRG can go to larger basis space). Shell model spaces for the VS-IMSRG Hamiltonian are protons and neutrons in $0p_{3/2,1/2}$ space for $^{6{\text -}12}\rm Be$, and protons in $0p_{3/2,1/2}$ and neutrons in $1s_{1/2}0d_{5/2,3/2}$ for $^{12{\text -}16}\rm Be$. A change in model space may bring some variation in calculated energy, e.g., a variation of about 1 MeV in $^{12}\rm Be$ with the two different model spaces above. This small uncertainty from different model spaces is common in VS-IMSRG calculations~\cite{S.R.Stroberg-Phys.Rev.Lett.126.022501(2021)}. We see that the angular momentum projection lowers the ground-state energies of $^{8{\text -}16}\rm Be$ by about 5-6 MeV, which makes calculated energies closer to data. The coupling to continuum can further lower the energy, which is more significant in nuclei near the dripline~\cite{Z.H.Sun-Phys.Lett.B.769.227(2017),B.S.Hu-Phys.Lett.B.802.135206(2020),B.S.Hu-PhysRevC.99.061302(2019),Y.Z.Ma-Phys.Lett.B.802.135257(2020),Y.Z.Ma-Phys.Lett.B.808.135673(2020), J.G.Li-PhysRevC.102.034302(2020)}. However, the inclusion of continuum partial waves increases dramatically the model dimension. D-IMSRG (with and without the projection) and VS-IMSRG calculations give the neutron dripline at $^{12}\rm Be$, while the experimental dripline position is at the next even-even isotope $^{14}\rm Be$. It is known that using different nuclear forces may lead to slightly different dripline positions. The continuum coupling can also affect the dripline position~\cite{Z.H.Sun-Phys.Lett.B.769.227(2017),B.S.Hu-Phys.Lett.B.802.135206(2020),B.S.Hu-PhysRevC.99.061302(2019),G.Hagen-Phys.Scr.91.063006(2016)}.

The D-IMSRG has also been applied to heavier nuclei of C, O, Ne and Mg isotopes, as shown in Fig. \ref{fig3} along with VS-IMSRG calculations and experimental data. The D-IMSRG calculations with the projection correction agree well with VS-IMSRG results and experimental data. For the closed-shell nuclei of $^{14}$C and $^{14,16,22,24,28}$O, we find that single-particle levels given by the $m$-scheme HF are degenerate with respect to the spin projection quantum number $m$, which indicates  spherical characteristic. Within the estimated uncertainty, the D-IMSRG calculation is identical to the VS-IMSRG result. However, for the expected closed-shell nuclei of $^{12,22}$C, the $m$-scheme HF gives nondegenerate single-particle levels with respect to $m$, indicating a deformation. The resulted angular momentum projection corrections are $-5.5$ and $-2.7$ MeV for the ground states of $^{12}$C and $^{22}$C, respectively, making the energies closer to VS-IMSRG results and data. For Ne and Mg isotopes near the neutron number $N =20$ island of inversion \cite{A.Poves-Phys.Lett.B.184.311(1987),E.K.Warburton-PhysRevC.41.1147(1990)}, the projection results in the energy gains of about 3-6 MeV. In the island-of-inversion nuclei, there is strong configuration mixing between $sd$ and $pf$ shells. The cross-shell mixing is missing in the present VS-IMSRG calculation, though the multishell VS-IMSRG has been proposed in Ref.~\cite{T.Miyagi-PhysRevC.102.034320(2020)}. In the D-IMSRG, the deformation effectively brings the deformation orbitals into the wave function of the state.

The charge radius is another important observable for nuclei. In the present work, we calculated the radii of the studied isotopes. The expectation value of the squared charge radius can be written as \cite{DCC-PhysRevC.102.051303(2020), G.Hagen-Nat.Phys.12.186(2016), B.S.Hu-PhysRevC.94.014303(2016)}:
\begin{equation}
 \langle R_{\rm ch}^2 \rangle=\langle R_p^2 \rangle+\langle r_p^2 \rangle+\frac{N}{Z}\langle r_n^2 \rangle+\langle r_{\rm DF}^2 \rangle+\langle r_{\rm so}^2 \rangle,
 \end{equation}
 where $R_p^2$ is the square of the intrinsic point-proton radius, and $r_{\rm so}^2$ is the spin-orbit correction, which can be calculated by the D-IMSRG. For other quantities in the equation, we usually take the proton radius squared $\langle r_p^2 \rangle=0.709 \ \rm fm^2$, the neutron radius squared $\langle r_n^2 \rangle=-0.106 \ \rm fm^2$, and the Darwin-Foldy term $\langle r_{\rm DF}^2 \rangle=\frac{3\hbar^2}{4m_{\rm p}^2c^2}=0.033 \ \rm fm^2$~\cite{DCC-PhysRevC.102.051303(2020), G.Hagen-Nat.Phys.12.186(2016), B.S.Hu-PhysRevC.94.014303(2016)}.

The convergence of the radius calculation shows a different trend with increasing the basis-space size, compared with the calculation of the ground-state energy, as discussed in \cite{S.K.Bogner-Nucl.Phys.A.801.21(2008),J.Hoppe-PhysRevC.100.024318(2019),T.Abe-PhysRevC.104.054315(2021)} and also found in the present D-IMSRG calculations. This indicates that the exponential fit is not applicable to the radius extrapolation, therefore no extrapolation has been attempted for the radius. In the D-IMSRG radius calculation, the angular momentum projection correction is also estimated using the HF wavefunction. The charge radii of Be, C, O, Ne and Mg isotopes have been investigated with a basis space of $N_{\rm shell}=10$, as shown in Fig.~\ref{fig4}, along with the VS-IMSRG calculations and experimental data available. The VS-IMSRG calculations were done with the $N_{\rm shell}=13$ basis space, and the valence spaces were chosen as the same as in the energy calculations above. For $^{20}\rm Ne$ and $^{34}\rm Mg$, our D-IMSRG charge radii are consistent with the recent projected CC calculations given in Ref.~\cite{PCC-arXiv.2201.07298(2022)}. The projection correction to the charge radius is small. The D-IMSRG radii with and without the projection correction are close to each other, and also well agree with the VS-IMSRG calculations except for $^8$Be in which the D-IMSRG radius is larger than the VS-IMSRG radius. This difference may be due to the large deformation of $^8$Be. In the calculation with the deformation degree of freedom, the deformation-sensitive orbitals descend and effectively enter the wave function of the deformed nucleus. The deformed basis states carry more correlations. In general, see Fig.~\ref{fig4}, the calculated charge radii by D-IMSRG and VS-IMSRG are reasonable compared with experiment data, although the $\rm NNLO_{\rm opt}$ interaction underestimates nuclear radii as commented in Ref. \cite{R.Kanungo-PhysRevLett.117.102501(2016)}.



In summary, starting from chiral interaction, we have developed an $ab \ initio$ deformed in-medium similarity renormalization group (D-IMSRG) in the deformed Hartree-Fock basis. The contribution of the angular momentum projection was estimated through the projected Hartree-Fock. Due to heavy computational cost, the D-IMSRG calculation was performed in a finite basis space, and the ground-state energy was extrapolated to the infinite basis space by an exponential fit to obtain the converged value. We investigated the ground-state energies of $^{8,10}\rm Be$, and compared with the NCSM and VS-IMSRG calculations. The D-IMSRG calculations with the angular momentum projection correction are in a good agreement with the NCSM results. We then calculated systematically the ground-state energies and charge radii of nuclei from light beryllium to medium-mass magnesium isotopes with the chiral $\rm NNLO_{\rm opt}$ interaction, giving reasonable results compared with the VS-IMSRG calculations and experiment data available. The D-IMSRG is a single-reference method, giving a straightforward calculation of open-shell nuclei. With the deformation, important deformed configurations can be efficiently included in the wave function, which makes the calculation more efficient for deformed nuclei.

\acknowledgments
Valuable discussions with Changfeng Jiao are gratefully acknowledged.
This work has been supported by the National Key Research and Development Program of China under Grant No. 2018YFA0404401; the National Natural Science Foundation of China under Grants No. 11835001, No. 11921006, No. 12035001 and No. 12105106; China Postdoctoral Science Foundation under Grant No. BX20200136; the State Key Laboratory of Nuclear Physics and Technology, Peking University under Grant No. NPT2020ZZ01; the U.S. Department
of Energy, Office of Science, Office of Nuclear Physics, under Award Nos.~DE-FG02-96ER40963 and DE-SC0018223; and the CUSTIPEN (China-U.S. Theory Institute for Physics with Exotic Nuclei) funded by The U.S. Department of Energy, Office of Science under Grant No. de-sc0009971. We acknowledge the High-Performance Computing Platform of Peking University for providing computational resources.

\bibliography{report}

\begin{thebibliography}{61}%
\makeatletter
\providecommand \@ifxundefined [1]{%
 \@ifx{#1\undefined}
}%
\providecommand \@ifnum [1]{%
 \ifnum #1\expandafter \@firstoftwo
 \else \expandafter \@secondoftwo
 \fi
}%
\providecommand \@ifx [1]{%
 \ifx #1\expandafter \@firstoftwo
 \else \expandafter \@secondoftwo
 \fi
}%
\providecommand \natexlab [1]{#1}%
\providecommand \enquote  [1]{``#1''}%
\providecommand \bibnamefont  [1]{#1}%
\providecommand \bibfnamefont [1]{#1}%
\providecommand \citenamefont [1]{#1}%
\providecommand \href@noop [0]{\@secondoftwo}%
\providecommand \href [0]{\begingroup \@sanitize@url \@href}%
\providecommand \@href[1]{\@@startlink{#1}\@@href}%
\providecommand \@@href[1]{\endgroup#1\@@endlink}%
\providecommand \@sanitize@url [0]{\catcode `\\12\catcode `\$12\catcode
  `\&12\catcode `\#12\catcode `\^12\catcode `\_12\catcode `\%12\relax}%
\providecommand \@@startlink[1]{}%
\providecommand \@@endlink[0]{}%
\providecommand \url  [0]{\begingroup\@sanitize@url \@url }%
\providecommand \@url [1]{\endgroup\@href {#1}{\urlprefix }}%
\providecommand \urlprefix  [0]{URL }%
\providecommand \Eprint [0]{\href }%
\providecommand \doibase [0]{http://dx.doi.org/}%
\providecommand \selectlanguage [0]{\@gobble}%
\providecommand \bibinfo  [0]{\@secondoftwo}%
\providecommand \bibfield  [0]{\@secondoftwo}%
\providecommand \translation [1]{[#1]}%
\providecommand \BibitemOpen [0]{}%
\providecommand \bibitemStop [0]{}%
\providecommand \bibitemNoStop [0]{.\EOS\space}%
\providecommand \EOS [0]{\spacefactor3000\relax}%
\providecommand \BibitemShut  [1]{\csname bibitem#1\endcsname}%
\let\auto@bib@innerbib\@empty
\bibitem [{\citenamefont {Navr\'atil}\ \emph {et~al.}(2000)\citenamefont
  {Navr\'atil}, \citenamefont {Vary},\ and\ \citenamefont
  {Barrett}}]{Navratil-PhysRevC.62.054311(2000)}%
  \BibitemOpen
  \bibfield  {author} {\bibinfo {author} {\bibfnamefont {P.}~\bibnamefont
  {Navr\'atil}}, \bibinfo {author} {\bibfnamefont {J.~P.}\ \bibnamefont
  {Vary}}, \ and\ \bibinfo {author} {\bibfnamefont {B.~R.}\ \bibnamefont
  {Barrett}},\ }\href {\doibase 10.1103/PhysRevC.62.054311} {\bibfield
  {journal} {\bibinfo  {journal} {Phys. Rev. C}\ }\textbf {\bibinfo {volume}
  {62}},\ \bibinfo {pages} {054311} (\bibinfo {year} {2000})}\BibitemShut
  {NoStop}%
\bibitem [{\citenamefont {Barrett}\ \emph {et~al.}(2013)\citenamefont
  {Barrett}, \citenamefont {Navrátil},\ and\ \citenamefont
  {Vary}}]{BARRETT2013131}%
  \BibitemOpen
  \bibfield  {author} {\bibinfo {author} {\bibfnamefont {B.~R.}\ \bibnamefont
  {Barrett}}, \bibinfo {author} {\bibfnamefont {P.}~\bibnamefont {Navrátil}},
  \ and\ \bibinfo {author} {\bibfnamefont {J.~P.}\ \bibnamefont {Vary}},\
  }\href {\doibase https://doi.org/10.1016/j.ppnp.2012.10.003} {\bibfield
  {journal} {\bibinfo  {journal} {Prog. Part. Nucl. Phys.}\ }\textbf {\bibinfo
  {volume} {69}},\ \bibinfo {pages} {131} (\bibinfo {year} {2013})}\BibitemShut
  {NoStop}%
\bibitem [{\citenamefont {Carlson}\ \emph {et~al.}(2015)\citenamefont
  {Carlson}, \citenamefont {Gandolfi}, \citenamefont {Pederiva}, \citenamefont
  {Pieper}, \citenamefont {Schiavilla}, \citenamefont {Schmidt},\ and\
  \citenamefont {Wiringa}}]{RevModPhys.87.1067}%
  \BibitemOpen
  \bibfield  {author} {\bibinfo {author} {\bibfnamefont {J.}~\bibnamefont
  {Carlson}}, \bibinfo {author} {\bibfnamefont {S.}~\bibnamefont {Gandolfi}},
  \bibinfo {author} {\bibfnamefont {F.}~\bibnamefont {Pederiva}}, \bibinfo
  {author} {\bibfnamefont {S.~C.}\ \bibnamefont {Pieper}}, \bibinfo {author}
  {\bibfnamefont {R.}~\bibnamefont {Schiavilla}}, \bibinfo {author}
  {\bibfnamefont {K.~E.}\ \bibnamefont {Schmidt}}, \ and\ \bibinfo {author}
  {\bibfnamefont {R.~B.}\ \bibnamefont {Wiringa}},\ }\href {\doibase
  10.1103/RevModPhys.87.1067} {\bibfield  {journal} {\bibinfo  {journal} {Rev.
  Mod. Phys.}\ }\textbf {\bibinfo {volume} {87}},\ \bibinfo {pages} {1067}
  (\bibinfo {year} {2015})}\BibitemShut {NoStop}%
\bibitem [{\citenamefont {Tsukiyama}\ \emph {et~al.}(2011)\citenamefont
  {Tsukiyama}, \citenamefont {Bogner},\ and\ \citenamefont
  {Schwenk}}]{IMSRG-PhysRevLett.106.222502(2011)}%
  \BibitemOpen
  \bibfield  {author} {\bibinfo {author} {\bibfnamefont {K.}~\bibnamefont
  {Tsukiyama}}, \bibinfo {author} {\bibfnamefont {S.~K.}\ \bibnamefont
  {Bogner}}, \ and\ \bibinfo {author} {\bibfnamefont {A.}~\bibnamefont
  {Schwenk}},\ }\href {\doibase 10.1103/PhysRevLett.106.222502} {\bibfield
  {journal} {\bibinfo  {journal} {Phys. Rev. Lett.}\ }\textbf {\bibinfo
  {volume} {106}},\ \bibinfo {pages} {222502} (\bibinfo {year}
  {2011})}\BibitemShut {NoStop}%
\bibitem [{\citenamefont {Hergert}\ \emph {et~al.}(2016)\citenamefont
  {Hergert}, \citenamefont {Bogner}, \citenamefont {Morris}, \citenamefont
  {Schwenk},\ and\ \citenamefont {Tsukiyama}}]{IMSRG-Phys.Rep.621.165(2016)}%
  \BibitemOpen
  \bibfield  {author} {\bibinfo {author} {\bibfnamefont {H.}~\bibnamefont
  {Hergert}}, \bibinfo {author} {\bibfnamefont {S.}~\bibnamefont {Bogner}},
  \bibinfo {author} {\bibfnamefont {T.}~\bibnamefont {Morris}}, \bibinfo
  {author} {\bibfnamefont {A.}~\bibnamefont {Schwenk}}, \ and\ \bibinfo
  {author} {\bibfnamefont {K.}~\bibnamefont {Tsukiyama}},\ }\href {\doibase
  https://doi.org/10.1016/j.physrep.2015.12.007} {\bibfield  {journal}
  {\bibinfo  {journal} {Phys. Rep.}\ }\textbf {\bibinfo {volume} {621}},\
  \bibinfo {pages} {165} (\bibinfo {year} {2016})}\BibitemShut {NoStop}%
\bibitem [{\citenamefont {Hagen}\ \emph {et~al.}(2014)\citenamefont {Hagen},
  \citenamefont {Papenbrock}, \citenamefont {Hjorth-Jensen},\ and\
  \citenamefont {Dean}}]{CC-Rep.Prog.Phys.77.096302(2014)}%
  \BibitemOpen
  \bibfield  {author} {\bibinfo {author} {\bibfnamefont {G.}~\bibnamefont
  {Hagen}}, \bibinfo {author} {\bibfnamefont {T.}~\bibnamefont {Papenbrock}},
  \bibinfo {author} {\bibfnamefont {M.}~\bibnamefont {Hjorth-Jensen}}, \ and\
  \bibinfo {author} {\bibfnamefont {D.~J.}\ \bibnamefont {Dean}},\ }\href
  {\doibase 10.1088/0034-4885/77/9/096302} {\bibfield  {journal} {\bibinfo
  {journal} {Rep. Prog. Phys.}\ }\textbf {\bibinfo {volume} {77}},\ \bibinfo
  {pages} {096302} (\bibinfo {year} {2014})}\BibitemShut {NoStop}%
\bibitem [{\citenamefont {Jansen}\ \emph {et~al.}(2014)\citenamefont {Jansen},
  \citenamefont {Engel}, \citenamefont {Hagen}, \citenamefont {Navratil},\ and\
  \citenamefont {Signoracci}}]{CC-PhysRevLett.113.142502(2014)}%
  \BibitemOpen
  \bibfield  {author} {\bibinfo {author} {\bibfnamefont {G.~R.}\ \bibnamefont
  {Jansen}}, \bibinfo {author} {\bibfnamefont {J.}~\bibnamefont {Engel}},
  \bibinfo {author} {\bibfnamefont {G.}~\bibnamefont {Hagen}}, \bibinfo
  {author} {\bibfnamefont {P.}~\bibnamefont {Navratil}}, \ and\ \bibinfo
  {author} {\bibfnamefont {A.}~\bibnamefont {Signoracci}},\ }\href {\doibase
  10.1103/PhysRevLett.113.142502} {\bibfield  {journal} {\bibinfo  {journal}
  {Phys. Rev. Lett.}\ }\textbf {\bibinfo {volume} {113}},\ \bibinfo {pages}
  {142502} (\bibinfo {year} {2014})}\BibitemShut {NoStop}%
\bibitem [{\citenamefont {Dickhoff}\ and\ \citenamefont
  {Barbieri}(2004)}]{W.H.Dickhoff-Prog.Part.Nucl.Phys.52.377(2004)}%
  \BibitemOpen
  \bibfield  {author} {\bibinfo {author} {\bibfnamefont {W.}~\bibnamefont
  {Dickhoff}}\ and\ \bibinfo {author} {\bibfnamefont {C.}~\bibnamefont
  {Barbieri}},\ }\href {\doibase https://doi.org/10.1016/j.ppnp.2004.02.038}
  {\bibfield  {journal} {\bibinfo  {journal} {Prog. Part. Nucl. Phys.}\
  }\textbf {\bibinfo {volume} {52}},\ \bibinfo {pages} {377} (\bibinfo {year}
  {2004})}\BibitemShut {NoStop}%
\bibitem [{\citenamefont {Som\`a}\ \emph {et~al.}(2013)\citenamefont {Som\`a},
  \citenamefont {Barbieri},\ and\ \citenamefont
  {Duguet}}]{V.Soma-PhysRevC.87.011303(2013)}%
  \BibitemOpen
  \bibfield  {author} {\bibinfo {author} {\bibfnamefont {V.}~\bibnamefont
  {Som\`a}}, \bibinfo {author} {\bibfnamefont {C.}~\bibnamefont {Barbieri}}, \
  and\ \bibinfo {author} {\bibfnamefont {T.}~\bibnamefont {Duguet}},\ }\href
  {\doibase 10.1103/PhysRevC.87.011303} {\bibfield  {journal} {\bibinfo
  {journal} {Phys. Rev. C}\ }\textbf {\bibinfo {volume} {87}},\ \bibinfo
  {pages} {011303} (\bibinfo {year} {2013})}\BibitemShut {NoStop}%
\bibitem [{\citenamefont {Tichai}\ \emph {et~al.}(2016)\citenamefont {Tichai},
  \citenamefont {Langhammer}, \citenamefont {Binder},\ and\ \citenamefont
  {Roth}}]{A.Tichai-Phys.Lett.B.756.283(2016)}%
  \BibitemOpen
  \bibfield  {author} {\bibinfo {author} {\bibfnamefont {A.}~\bibnamefont
  {Tichai}}, \bibinfo {author} {\bibfnamefont {J.}~\bibnamefont {Langhammer}},
  \bibinfo {author} {\bibfnamefont {S.}~\bibnamefont {Binder}}, \ and\ \bibinfo
  {author} {\bibfnamefont {R.}~\bibnamefont {Roth}},\ }\href {\doibase
  https://doi.org/10.1016/j.physletb.2016.03.029} {\bibfield  {journal}
  {\bibinfo  {journal} {Phys. Lett. B}\ }\textbf {\bibinfo {volume} {756}},\
  \bibinfo {pages} {283} (\bibinfo {year} {2016})}\BibitemShut {NoStop}%
\bibitem [{\citenamefont {Hu}\ \emph {et~al.}(2016)\citenamefont {Hu},
  \citenamefont {Xu}, \citenamefont {Sun}, \citenamefont {Vary},\ and\
  \citenamefont {Li}}]{B.S.Hu-PhysRevC.94.014303(2016)}%
  \BibitemOpen
  \bibfield  {author} {\bibinfo {author} {\bibfnamefont {B.~S.}\ \bibnamefont
  {Hu}}, \bibinfo {author} {\bibfnamefont {F.~R.}\ \bibnamefont {Xu}}, \bibinfo
  {author} {\bibfnamefont {Z.~H.}\ \bibnamefont {Sun}}, \bibinfo {author}
  {\bibfnamefont {J.~P.}\ \bibnamefont {Vary}}, \ and\ \bibinfo {author}
  {\bibfnamefont {T.}~\bibnamefont {Li}},\ }\href {\doibase
  10.1103/PhysRevC.94.014303} {\bibfield  {journal} {\bibinfo  {journal} {Phys.
  Rev. C}\ }\textbf {\bibinfo {volume} {94}},\ \bibinfo {pages} {014303}
  (\bibinfo {year} {2016})}\BibitemShut {NoStop}%
\bibitem [{\citenamefont {Tichai}\ \emph {et~al.}(2018)\citenamefont {Tichai},
  \citenamefont {Arthuis}, \citenamefont {Duguet}, \citenamefont {Hergert},
  \citenamefont {Somà},\ and\ \citenamefont
  {Roth}}]{A.Tichai-Phys.Lett.B.786.195(2018)}%
  \BibitemOpen
  \bibfield  {author} {\bibinfo {author} {\bibfnamefont {A.}~\bibnamefont
  {Tichai}}, \bibinfo {author} {\bibfnamefont {P.}~\bibnamefont {Arthuis}},
  \bibinfo {author} {\bibfnamefont {T.}~\bibnamefont {Duguet}}, \bibinfo
  {author} {\bibfnamefont {H.}~\bibnamefont {Hergert}}, \bibinfo {author}
  {\bibfnamefont {V.}~\bibnamefont {Somà}}, \ and\ \bibinfo {author}
  {\bibfnamefont {R.}~\bibnamefont {Roth}},\ }\href {\doibase
  https://doi.org/10.1016/j.physletb.2018.09.044} {\bibfield  {journal}
  {\bibinfo  {journal} {Phys. Lett. B}\ }\textbf {\bibinfo {volume} {786}},\
  \bibinfo {pages} {195} (\bibinfo {year} {2018})}\BibitemShut {NoStop}%
\bibitem [{\citenamefont {Hagen}\ \emph {et~al.}(2008)\citenamefont {Hagen},
  \citenamefont {Papenbrock}, \citenamefont {Dean},\ and\ \citenamefont
  {Hjorth-Jensen}}]{CC-PhysRevLett.101.092502(2008)}%
  \BibitemOpen
  \bibfield  {author} {\bibinfo {author} {\bibfnamefont {G.}~\bibnamefont
  {Hagen}}, \bibinfo {author} {\bibfnamefont {T.}~\bibnamefont {Papenbrock}},
  \bibinfo {author} {\bibfnamefont {D.~J.}\ \bibnamefont {Dean}}, \ and\
  \bibinfo {author} {\bibfnamefont {M.}~\bibnamefont {Hjorth-Jensen}},\ }\href
  {\doibase 10.1103/PhysRevLett.101.092502} {\bibfield  {journal} {\bibinfo
  {journal} {Phys. Rev. Lett.}\ }\textbf {\bibinfo {volume} {101}},\ \bibinfo
  {pages} {092502} (\bibinfo {year} {2008})}\BibitemShut {NoStop}%
\bibitem [{\citenamefont {Hergert}\ \emph {et~al.}(2013)\citenamefont
  {Hergert}, \citenamefont {Binder}, \citenamefont {Calci}, \citenamefont
  {Langhammer},\ and\ \citenamefont
  {Roth}}]{MRIMSRG-PhysRevLett.110.242501(2013)}%
  \BibitemOpen
  \bibfield  {author} {\bibinfo {author} {\bibfnamefont {H.}~\bibnamefont
  {Hergert}}, \bibinfo {author} {\bibfnamefont {S.}~\bibnamefont {Binder}},
  \bibinfo {author} {\bibfnamefont {A.}~\bibnamefont {Calci}}, \bibinfo
  {author} {\bibfnamefont {J.}~\bibnamefont {Langhammer}}, \ and\ \bibinfo
  {author} {\bibfnamefont {R.}~\bibnamefont {Roth}},\ }\href {\doibase
  10.1103/PhysRevLett.110.242501} {\bibfield  {journal} {\bibinfo  {journal}
  {Phys. Rev. Lett.}\ }\textbf {\bibinfo {volume} {110}},\ \bibinfo {pages}
  {242501} (\bibinfo {year} {2013})}\BibitemShut {NoStop}%
\bibitem [{\citenamefont {Hergert}\ \emph {et~al.}(2014)\citenamefont
  {Hergert}, \citenamefont {Bogner}, \citenamefont {Morris}, \citenamefont
  {Binder}, \citenamefont {Calci}, \citenamefont {Langhammer},\ and\
  \citenamefont {Roth}}]{MRIMSRG-PhysRevC.90.041302(2014)}%
  \BibitemOpen
  \bibfield  {author} {\bibinfo {author} {\bibfnamefont {H.}~\bibnamefont
  {Hergert}}, \bibinfo {author} {\bibfnamefont {S.~K.}\ \bibnamefont {Bogner}},
  \bibinfo {author} {\bibfnamefont {T.~D.}\ \bibnamefont {Morris}}, \bibinfo
  {author} {\bibfnamefont {S.}~\bibnamefont {Binder}}, \bibinfo {author}
  {\bibfnamefont {A.}~\bibnamefont {Calci}}, \bibinfo {author} {\bibfnamefont
  {J.}~\bibnamefont {Langhammer}}, \ and\ \bibinfo {author} {\bibfnamefont
  {R.}~\bibnamefont {Roth}},\ }\href {\doibase 10.1103/PhysRevC.90.041302}
  {\bibfield  {journal} {\bibinfo  {journal} {Phys. Rev. C}\ }\textbf {\bibinfo
  {volume} {90}},\ \bibinfo {pages} {041302} (\bibinfo {year}
  {2014})}\BibitemShut {NoStop}%
\bibitem [{\citenamefont {Signoracci}\ \emph {et~al.}(2015)\citenamefont
  {Signoracci}, \citenamefont {Duguet}, \citenamefont {Hagen},\ and\
  \citenamefont {Jansen}}]{HFBCC-PhysRevC.91.064320(2015)}%
  \BibitemOpen
  \bibfield  {author} {\bibinfo {author} {\bibfnamefont {A.}~\bibnamefont
  {Signoracci}}, \bibinfo {author} {\bibfnamefont {T.}~\bibnamefont {Duguet}},
  \bibinfo {author} {\bibfnamefont {G.}~\bibnamefont {Hagen}}, \ and\ \bibinfo
  {author} {\bibfnamefont {G.~R.}\ \bibnamefont {Jansen}},\ }\href {\doibase
  10.1103/PhysRevC.91.064320} {\bibfield  {journal} {\bibinfo  {journal} {Phys.
  Rev. C}\ }\textbf {\bibinfo {volume} {91}},\ \bibinfo {pages} {064320}
  (\bibinfo {year} {2015})}\BibitemShut {NoStop}%
\bibitem [{\citenamefont {Frosini}\ \emph
  {et~al.}(2022{\natexlab{a}})\citenamefont {Frosini}, \citenamefont {Duguet},
  \citenamefont {Ebran},\ and\ \citenamefont {Somà}}]{mMBPT1}%
  \BibitemOpen
  \bibfield  {author} {\bibinfo {author} {\bibfnamefont {M.}~\bibnamefont
  {Frosini}}, \bibinfo {author} {\bibfnamefont {T.}~\bibnamefont {Duguet}},
  \bibinfo {author} {\bibfnamefont {J.-P.}\ \bibnamefont {Ebran}}, \ and\
  \bibinfo {author} {\bibfnamefont {V.}~\bibnamefont {Somà}},\ }\href@noop {}
  {} (\bibinfo {year} {2022}{\natexlab{a}}),\ \Eprint
  {http://arxiv.org/abs/2110.15737} {arXiv:2110.15737 [nucl-th]} \BibitemShut
  {NoStop}%
\bibitem [{\citenamefont {Frosini}\ \emph
  {et~al.}(2022{\natexlab{b}})\citenamefont {Frosini}, \citenamefont {Duguet},
  \citenamefont {Ebran}, \citenamefont {Bally}, \citenamefont {Mongelli},
  \citenamefont {Rodríguez}, \citenamefont {Roth},\ and\ \citenamefont
  {Somà}}]{mMBPT2}%
  \BibitemOpen
  \bibfield  {author} {\bibinfo {author} {\bibfnamefont {M.}~\bibnamefont
  {Frosini}}, \bibinfo {author} {\bibfnamefont {T.}~\bibnamefont {Duguet}},
  \bibinfo {author} {\bibfnamefont {J.-P.}\ \bibnamefont {Ebran}}, \bibinfo
  {author} {\bibfnamefont {B.}~\bibnamefont {Bally}}, \bibinfo {author}
  {\bibfnamefont {T.}~\bibnamefont {Mongelli}}, \bibinfo {author}
  {\bibfnamefont {T.~R.}\ \bibnamefont {Rodríguez}}, \bibinfo {author}
  {\bibfnamefont {R.}~\bibnamefont {Roth}}, \ and\ \bibinfo {author}
  {\bibfnamefont {V.}~\bibnamefont {Somà}},\ }\href@noop {} {} (\bibinfo
  {year} {2022}{\natexlab{b}}),\ \Eprint {http://arxiv.org/abs/2111.00797}
  {arXiv:2111.00797 [nucl-th]} \BibitemShut {NoStop}%
\bibitem [{\citenamefont {Frosini}\ \emph
  {et~al.}(2022{\natexlab{c}})\citenamefont {Frosini}, \citenamefont {Duguet},
  \citenamefont {Ebran}, \citenamefont {Bally}, \citenamefont {Hergert},
  \citenamefont {Rodríguez}, \citenamefont {Roth}, \citenamefont {Yao},\ and\
  \citenamefont {Somà}}]{mMBPT3}%
  \BibitemOpen
  \bibfield  {author} {\bibinfo {author} {\bibfnamefont {M.}~\bibnamefont
  {Frosini}}, \bibinfo {author} {\bibfnamefont {T.}~\bibnamefont {Duguet}},
  \bibinfo {author} {\bibfnamefont {J.-P.}\ \bibnamefont {Ebran}}, \bibinfo
  {author} {\bibfnamefont {B.}~\bibnamefont {Bally}}, \bibinfo {author}
  {\bibfnamefont {H.}~\bibnamefont {Hergert}}, \bibinfo {author} {\bibfnamefont
  {T.~R.}\ \bibnamefont {Rodríguez}}, \bibinfo {author} {\bibfnamefont
  {R.}~\bibnamefont {Roth}}, \bibinfo {author} {\bibfnamefont {J.}~\bibnamefont
  {Yao}}, \ and\ \bibinfo {author} {\bibfnamefont {V.}~\bibnamefont {Somà}},\
  }\href@noop {} {} (\bibinfo {year} {2022}{\natexlab{c}}),\ \Eprint
  {http://arxiv.org/abs/2111.01461} {arXiv:2111.01461 [nucl-th]} \BibitemShut
  {NoStop}%
\bibitem [{\citenamefont {Bogner}\ \emph {et~al.}(2014)\citenamefont {Bogner},
  \citenamefont {Hergert}, \citenamefont {Holt}, \citenamefont {Schwenk},
  \citenamefont {Binder}, \citenamefont {Calci}, \citenamefont {Langhammer},\
  and\ \citenamefont {Roth}}]{IMSRG-PhysRevLett.113.142501(2014)}%
  \BibitemOpen
  \bibfield  {author} {\bibinfo {author} {\bibfnamefont {S.~K.}\ \bibnamefont
  {Bogner}}, \bibinfo {author} {\bibfnamefont {H.}~\bibnamefont {Hergert}},
  \bibinfo {author} {\bibfnamefont {J.~D.}\ \bibnamefont {Holt}}, \bibinfo
  {author} {\bibfnamefont {A.}~\bibnamefont {Schwenk}}, \bibinfo {author}
  {\bibfnamefont {S.}~\bibnamefont {Binder}}, \bibinfo {author} {\bibfnamefont
  {A.}~\bibnamefont {Calci}}, \bibinfo {author} {\bibfnamefont
  {J.}~\bibnamefont {Langhammer}}, \ and\ \bibinfo {author} {\bibfnamefont
  {R.}~\bibnamefont {Roth}},\ }\href {\doibase 10.1103/PhysRevLett.113.142501}
  {\bibfield  {journal} {\bibinfo  {journal} {Phys. Rev. Lett.}\ }\textbf
  {\bibinfo {volume} {113}},\ \bibinfo {pages} {142501} (\bibinfo {year}
  {2014})}\BibitemShut {NoStop}%
\bibitem [{\citenamefont {Stroberg}\ \emph {et~al.}(2017)\citenamefont
  {Stroberg}, \citenamefont {Calci}, \citenamefont {Hergert}, \citenamefont
  {Holt}, \citenamefont {Bogner}, \citenamefont {Roth},\ and\ \citenamefont
  {Schwenk}}]{IMSRG-PhysRevLett.118.032502(2017)}%
  \BibitemOpen
  \bibfield  {author} {\bibinfo {author} {\bibfnamefont {S.~R.}\ \bibnamefont
  {Stroberg}}, \bibinfo {author} {\bibfnamefont {A.}~\bibnamefont {Calci}},
  \bibinfo {author} {\bibfnamefont {H.}~\bibnamefont {Hergert}}, \bibinfo
  {author} {\bibfnamefont {J.~D.}\ \bibnamefont {Holt}}, \bibinfo {author}
  {\bibfnamefont {S.~K.}\ \bibnamefont {Bogner}}, \bibinfo {author}
  {\bibfnamefont {R.}~\bibnamefont {Roth}}, \ and\ \bibinfo {author}
  {\bibfnamefont {A.}~\bibnamefont {Schwenk}},\ }\href {\doibase
  10.1103/PhysRevLett.118.032502} {\bibfield  {journal} {\bibinfo  {journal}
  {Phys. Rev. Lett.}\ }\textbf {\bibinfo {volume} {118}},\ \bibinfo {pages}
  {032502} (\bibinfo {year} {2017})}\BibitemShut {NoStop}%
\bibitem [{\citenamefont {Sun}\ \emph {et~al.}(2018)\citenamefont {Sun},
  \citenamefont {Morris}, \citenamefont {Hagen}, \citenamefont {Jansen},\ and\
  \citenamefont {Papenbrock}}]{CC-PhysRevC.98.054320(2018)}%
  \BibitemOpen
  \bibfield  {author} {\bibinfo {author} {\bibfnamefont {Z.~H.}\ \bibnamefont
  {Sun}}, \bibinfo {author} {\bibfnamefont {T.~D.}\ \bibnamefont {Morris}},
  \bibinfo {author} {\bibfnamefont {G.}~\bibnamefont {Hagen}}, \bibinfo
  {author} {\bibfnamefont {G.~R.}\ \bibnamefont {Jansen}}, \ and\ \bibinfo
  {author} {\bibfnamefont {T.}~\bibnamefont {Papenbrock}},\ }\href {\doibase
  10.1103/PhysRevC.98.054320} {\bibfield  {journal} {\bibinfo  {journal} {Phys.
  Rev. C}\ }\textbf {\bibinfo {volume} {98}},\ \bibinfo {pages} {054320}
  (\bibinfo {year} {2018})}\BibitemShut {NoStop}%
\bibitem [{\citenamefont {Hjorth-Jensen}\ \emph {et~al.}(1995)\citenamefont
  {Hjorth-Jensen}, \citenamefont {Kuo},\ and\ \citenamefont
  {Osnes}}]{M.Hjorth.Jensen-Phys.Rep.261.125(1995)}%
  \BibitemOpen
  \bibfield  {author} {\bibinfo {author} {\bibfnamefont {M.}~\bibnamefont
  {Hjorth-Jensen}}, \bibinfo {author} {\bibfnamefont {T.~T.}\ \bibnamefont
  {Kuo}}, \ and\ \bibinfo {author} {\bibfnamefont {E.}~\bibnamefont {Osnes}},\
  }\href {\doibase https://doi.org/10.1016/0370-1573(95)00012-6} {\bibfield
  {journal} {\bibinfo  {journal} {Phys. Rep.}\ }\textbf {\bibinfo {volume}
  {261}},\ \bibinfo {pages} {125} (\bibinfo {year} {1995})}\BibitemShut
  {NoStop}%
\bibitem [{\citenamefont {Coraggio}\ \emph {et~al.}(2012)\citenamefont
  {Coraggio}, \citenamefont {Covello}, \citenamefont {Gargano}, \citenamefont
  {Itaco},\ and\ \citenamefont {Kuo}}]{L.Coraggio-Ann.Phys.327.2125(2012)}%
  \BibitemOpen
  \bibfield  {author} {\bibinfo {author} {\bibfnamefont {L.}~\bibnamefont
  {Coraggio}}, \bibinfo {author} {\bibfnamefont {A.}~\bibnamefont {Covello}},
  \bibinfo {author} {\bibfnamefont {A.}~\bibnamefont {Gargano}}, \bibinfo
  {author} {\bibfnamefont {N.}~\bibnamefont {Itaco}}, \ and\ \bibinfo {author}
  {\bibfnamefont {T.}~\bibnamefont {Kuo}},\ }\href {\doibase
  https://doi.org/10.1016/j.aop.2012.04.013} {\bibfield  {journal} {\bibinfo
  {journal} {Ann. Phys.}\ }\textbf {\bibinfo {volume} {327}},\ \bibinfo {pages}
  {2125} (\bibinfo {year} {2012})}\BibitemShut {NoStop}%
\bibitem [{\citenamefont {Sun}\ \emph {et~al.}(2017)\citenamefont {Sun},
  \citenamefont {Wu}, \citenamefont {Zhao}, \citenamefont {Hu}, \citenamefont
  {Dai},\ and\ \citenamefont {Xu}}]{Z.H.Sun-Phys.Lett.B.769.227(2017)}%
  \BibitemOpen
  \bibfield  {author} {\bibinfo {author} {\bibfnamefont {Z.~H.}\ \bibnamefont
  {Sun}}, \bibinfo {author} {\bibfnamefont {Q.}~\bibnamefont {Wu}}, \bibinfo
  {author} {\bibfnamefont {Z.~H.}\ \bibnamefont {Zhao}}, \bibinfo {author}
  {\bibfnamefont {B.~S.}\ \bibnamefont {Hu}}, \bibinfo {author} {\bibfnamefont
  {S.~J.}\ \bibnamefont {Dai}}, \ and\ \bibinfo {author} {\bibfnamefont
  {F.~R.}\ \bibnamefont {Xu}},\ }\href {\doibase
  https://doi.org/10.1016/j.physletb.2017.03.054} {\bibfield  {journal}
  {\bibinfo  {journal} {Phys. Lett. B}\ }\textbf {\bibinfo {volume} {769}},\
  \bibinfo {pages} {227} (\bibinfo {year} {2017})}\BibitemShut {NoStop}%
\bibitem [{\citenamefont {Hu}\ \emph {et~al.}(2020)\citenamefont {Hu},
  \citenamefont {Wu}, \citenamefont {Li}, \citenamefont {Ma}, \citenamefont
  {Sun}, \citenamefont {Michel},\ and\ \citenamefont
  {Xu}}]{B.S.Hu-Phys.Lett.B.802.135206(2020)}%
  \BibitemOpen
  \bibfield  {author} {\bibinfo {author} {\bibfnamefont {B.~S.}\ \bibnamefont
  {Hu}}, \bibinfo {author} {\bibfnamefont {Q.}~\bibnamefont {Wu}}, \bibinfo
  {author} {\bibfnamefont {J.~G.}\ \bibnamefont {Li}}, \bibinfo {author}
  {\bibfnamefont {Y.~Z.}\ \bibnamefont {Ma}}, \bibinfo {author} {\bibfnamefont
  {Z.~H.}\ \bibnamefont {Sun}}, \bibinfo {author} {\bibfnamefont
  {N.}~\bibnamefont {Michel}}, \ and\ \bibinfo {author} {\bibfnamefont {F.~R.}\
  \bibnamefont {Xu}},\ }\href {\doibase
  https://doi.org/10.1016/j.physletb.2020.135206} {\bibfield  {journal}
  {\bibinfo  {journal} {Phys. Lett. B}\ }\textbf {\bibinfo {volume} {802}},\
  \bibinfo {pages} {135206} (\bibinfo {year} {2020})}\BibitemShut {NoStop}%
\bibitem [{\citenamefont {Duguet}(2015)}]{Duguet_2014}%
  \BibitemOpen
  \bibfield  {author} {\bibinfo {author} {\bibfnamefont {T.}~\bibnamefont
  {Duguet}},\ }\href {\doibase 10.1088/0954-3899/42/2/025107} {\bibfield
  {journal} {\bibinfo  {journal} {J. Phys. G.}\ }\textbf {\bibinfo {volume}
  {42}},\ \bibinfo {pages} {025107} (\bibinfo {year} {2015})}\BibitemShut
  {NoStop}%
\bibitem [{\citenamefont {Novario}\ \emph {et~al.}(2020)\citenamefont
  {Novario}, \citenamefont {Hagen}, \citenamefont {Jansen},\ and\ \citenamefont
  {Papenbrock}}]{DCC-PhysRevC.102.051303(2020)}%
  \BibitemOpen
  \bibfield  {author} {\bibinfo {author} {\bibfnamefont {S.~J.}\ \bibnamefont
  {Novario}}, \bibinfo {author} {\bibfnamefont {G.}~\bibnamefont {Hagen}},
  \bibinfo {author} {\bibfnamefont {G.~R.}\ \bibnamefont {Jansen}}, \ and\
  \bibinfo {author} {\bibfnamefont {T.}~\bibnamefont {Papenbrock}},\ }\href
  {\doibase 10.1103/PhysRevC.102.051303} {\bibfield  {journal} {\bibinfo
  {journal} {Phys. Rev. C}\ }\textbf {\bibinfo {volume} {102}},\ \bibinfo
  {pages} {051303} (\bibinfo {year} {2020})}\BibitemShut {NoStop}%
\bibitem [{\citenamefont {Hagen}\ \emph {et~al.}(2022)\citenamefont {Hagen},
  \citenamefont {Novario}, \citenamefont {Sun}, \citenamefont {Papenbrock},
  \citenamefont {Jansen}, \citenamefont {Lietz}, \citenamefont {Duguet},\ and\
  \citenamefont {Tichai}}]{PCC-arXiv.2201.07298(2022)}%
  \BibitemOpen
  \bibfield  {author} {\bibinfo {author} {\bibfnamefont {G.}~\bibnamefont
  {Hagen}}, \bibinfo {author} {\bibfnamefont {S.~J.}\ \bibnamefont {Novario}},
  \bibinfo {author} {\bibfnamefont {Z.~H.}\ \bibnamefont {Sun}}, \bibinfo
  {author} {\bibfnamefont {T.}~\bibnamefont {Papenbrock}}, \bibinfo {author}
  {\bibfnamefont {G.~R.}\ \bibnamefont {Jansen}}, \bibinfo {author}
  {\bibfnamefont {J.~G.}\ \bibnamefont {Lietz}}, \bibinfo {author}
  {\bibfnamefont {T.}~\bibnamefont {Duguet}}, \ and\ \bibinfo {author}
  {\bibfnamefont {A.}~\bibnamefont {Tichai}},\ }\href@noop {} {} (\bibinfo
  {year} {2022}),\ \Eprint {http://arxiv.org/abs/2201.07298} {arXiv:2201.07298
  [nucl-th]} \BibitemShut {NoStop}%
\bibitem [{\citenamefont {Novario}\ \emph {et~al.}(2021)\citenamefont
  {Novario}, \citenamefont {Gysbers}, \citenamefont {Engel}, \citenamefont
  {Hagen}, \citenamefont {Jansen}, \citenamefont {Morris}, \citenamefont
  {Navr\'atil}, \citenamefont {Papenbrock},\ and\ \citenamefont
  {Quaglioni}}]{DCC-PhysRevLett.126.182502(2021)}%
  \BibitemOpen
  \bibfield  {author} {\bibinfo {author} {\bibfnamefont {S.}~\bibnamefont
  {Novario}}, \bibinfo {author} {\bibfnamefont {P.}~\bibnamefont {Gysbers}},
  \bibinfo {author} {\bibfnamefont {J.}~\bibnamefont {Engel}}, \bibinfo
  {author} {\bibfnamefont {G.}~\bibnamefont {Hagen}}, \bibinfo {author}
  {\bibfnamefont {G.~R.}\ \bibnamefont {Jansen}}, \bibinfo {author}
  {\bibfnamefont {T.~D.}\ \bibnamefont {Morris}}, \bibinfo {author}
  {\bibfnamefont {P.}~\bibnamefont {Navr\'atil}}, \bibinfo {author}
  {\bibfnamefont {T.}~\bibnamefont {Papenbrock}}, \ and\ \bibinfo {author}
  {\bibfnamefont {S.}~\bibnamefont {Quaglioni}},\ }\href {\doibase
  10.1103/PhysRevLett.126.182502} {\bibfield  {journal} {\bibinfo  {journal}
  {Phys. Rev. Lett.}\ }\textbf {\bibinfo {volume} {126}},\ \bibinfo {pages}
  {182502} (\bibinfo {year} {2021})}\BibitemShut {NoStop}%
\bibitem [{\citenamefont {Ring}\ and\ \citenamefont {Schuck}()}]{book}%
  \BibitemOpen
  \bibfield  {author} {\bibinfo {author} {\bibfnamefont {P.}~\bibnamefont
  {Ring}}\ and\ \bibinfo {author} {\bibfnamefont {P.}~\bibnamefont {Schuck}},\
  }\href@noop {} {\enquote {\bibinfo {title} {The nuclear many-body
  problems},}\ }\bibinfo {note} {(Springer, Heidelberg, 1980)}\BibitemShut
  {NoStop}%
\bibitem [{\citenamefont {Dytrych}\ \emph {et~al.}(2013)\citenamefont
  {Dytrych}, \citenamefont {Launey}, \citenamefont {Draayer}, \citenamefont
  {Maris}, \citenamefont {Vary}, \citenamefont {Saule}, \citenamefont
  {Catalyurek}, \citenamefont {Sosonkina}, \citenamefont {Langr},\ and\
  \citenamefont {Caprio}}]{T.Dytrych-PhysRevLett.111.252501(2013)}%
  \BibitemOpen
  \bibfield  {author} {\bibinfo {author} {\bibfnamefont {T.}~\bibnamefont
  {Dytrych}}, \bibinfo {author} {\bibfnamefont {K.~D.}\ \bibnamefont {Launey}},
  \bibinfo {author} {\bibfnamefont {J.~P.}\ \bibnamefont {Draayer}}, \bibinfo
  {author} {\bibfnamefont {P.}~\bibnamefont {Maris}}, \bibinfo {author}
  {\bibfnamefont {J.~P.}\ \bibnamefont {Vary}}, \bibinfo {author}
  {\bibfnamefont {E.}~\bibnamefont {Saule}}, \bibinfo {author} {\bibfnamefont
  {U.}~\bibnamefont {Catalyurek}}, \bibinfo {author} {\bibfnamefont
  {M.}~\bibnamefont {Sosonkina}}, \bibinfo {author} {\bibfnamefont
  {D.}~\bibnamefont {Langr}}, \ and\ \bibinfo {author} {\bibfnamefont {M.~A.}\
  \bibnamefont {Caprio}},\ }\href {\doibase 10.1103/PhysRevLett.111.252501}
  {\bibfield  {journal} {\bibinfo  {journal} {Phys. Rev. Lett.}\ }\textbf
  {\bibinfo {volume} {111}},\ \bibinfo {pages} {252501} (\bibinfo {year}
  {2013})}\BibitemShut {NoStop}%
\bibitem [{\citenamefont {Dytrych}\ \emph {et~al.}(2020)\citenamefont
  {Dytrych}, \citenamefont {Launey}, \citenamefont {Draayer}, \citenamefont
  {Rowe}, \citenamefont {Wood}, \citenamefont {Rosensteel}, \citenamefont
  {Bahri}, \citenamefont {Langr},\ and\ \citenamefont
  {Baker}}]{T.Dytrych-PhysRevLett.124.042501(2020)}%
  \BibitemOpen
  \bibfield  {author} {\bibinfo {author} {\bibfnamefont {T.}~\bibnamefont
  {Dytrych}}, \bibinfo {author} {\bibfnamefont {K.~D.}\ \bibnamefont {Launey}},
  \bibinfo {author} {\bibfnamefont {J.~P.}\ \bibnamefont {Draayer}}, \bibinfo
  {author} {\bibfnamefont {D.~J.}\ \bibnamefont {Rowe}}, \bibinfo {author}
  {\bibfnamefont {J.~L.}\ \bibnamefont {Wood}}, \bibinfo {author}
  {\bibfnamefont {G.}~\bibnamefont {Rosensteel}}, \bibinfo {author}
  {\bibfnamefont {C.}~\bibnamefont {Bahri}}, \bibinfo {author} {\bibfnamefont
  {D.}~\bibnamefont {Langr}}, \ and\ \bibinfo {author} {\bibfnamefont {R.~B.}\
  \bibnamefont {Baker}},\ }\href {\doibase 10.1103/PhysRevLett.124.042501}
  {\bibfield  {journal} {\bibinfo  {journal} {Phys. Rev. Lett.}\ }\textbf
  {\bibinfo {volume} {124}},\ \bibinfo {pages} {042501} (\bibinfo {year}
  {2020})}\BibitemShut {NoStop}%
\bibitem [{\citenamefont {Ekstr\"om}\ \emph {et~al.}(2013)\citenamefont
  {Ekstr\"om}, \citenamefont {Baardsen}, \citenamefont {Forss\'en},
  \citenamefont {Hagen}, \citenamefont {Hjorth-Jensen}, \citenamefont {Jansen},
  \citenamefont {Machleidt}, \citenamefont {Nazarewicz}, \citenamefont
  {Papenbrock}, \citenamefont {Sarich},\ and\ \citenamefont
  {Wild}}]{A.Ekstrom-PhysRevLett.110.192502(2013)}%
  \BibitemOpen
  \bibfield  {author} {\bibinfo {author} {\bibfnamefont {A.}~\bibnamefont
  {Ekstr\"om}}, \bibinfo {author} {\bibfnamefont {G.}~\bibnamefont {Baardsen}},
  \bibinfo {author} {\bibfnamefont {C.}~\bibnamefont {Forss\'en}}, \bibinfo
  {author} {\bibfnamefont {G.}~\bibnamefont {Hagen}}, \bibinfo {author}
  {\bibfnamefont {M.}~\bibnamefont {Hjorth-Jensen}}, \bibinfo {author}
  {\bibfnamefont {G.~R.}\ \bibnamefont {Jansen}}, \bibinfo {author}
  {\bibfnamefont {R.}~\bibnamefont {Machleidt}}, \bibinfo {author}
  {\bibfnamefont {W.}~\bibnamefont {Nazarewicz}}, \bibinfo {author}
  {\bibfnamefont {T.}~\bibnamefont {Papenbrock}}, \bibinfo {author}
  {\bibfnamefont {J.}~\bibnamefont {Sarich}}, \ and\ \bibinfo {author}
  {\bibfnamefont {S.~M.}\ \bibnamefont {Wild}},\ }\href {\doibase
  10.1103/PhysRevLett.110.192502} {\bibfield  {journal} {\bibinfo  {journal}
  {Phys. Rev. Lett.}\ }\textbf {\bibinfo {volume} {110}},\ \bibinfo {pages}
  {192502} (\bibinfo {year} {2013})}\BibitemShut {NoStop}%
\bibitem [{\citenamefont {Kanungo}\ \emph {et~al.}(2016)\citenamefont
  {Kanungo}, \citenamefont {Horiuchi}, \citenamefont {Hagen}, \citenamefont
  {Jansen}, \citenamefont {Navratil}, \citenamefont {Ameil}, \citenamefont
  {Atkinson}, \citenamefont {Ayyad}, \citenamefont {Cortina-Gil}, \citenamefont
  {Dillmann}, \citenamefont {Estrad\'e}, \citenamefont {Evdokimov},
  \citenamefont {Farinon}, \citenamefont {Geissel}, \citenamefont {Guastalla},
  \citenamefont {Janik}, \citenamefont {Kimura}, \citenamefont {Kn\"obel},
  \citenamefont {Kurcewicz}, \citenamefont {Litvinov}, \citenamefont {Marta},
  \citenamefont {Mostazo}, \citenamefont {Mukha}, \citenamefont {Nociforo},
  \citenamefont {Ong}, \citenamefont {Pietri}, \citenamefont {Prochazka},
  \citenamefont {Scheidenberger}, \citenamefont {Sitar}, \citenamefont
  {Strmen}, \citenamefont {Suzuki}, \citenamefont {Takechi}, \citenamefont
  {Tanaka}, \citenamefont {Tanihata}, \citenamefont {Terashima}, \citenamefont
  {Vargas}, \citenamefont {Weick},\ and\ \citenamefont
  {Winfield}}]{R.Kanungo-PhysRevLett.117.102501(2016)}%
  \BibitemOpen
  \bibfield  {author} {\bibinfo {author} {\bibfnamefont {R.}~\bibnamefont
  {Kanungo}}, \bibinfo {author} {\bibfnamefont {W.}~\bibnamefont {Horiuchi}},
  \bibinfo {author} {\bibfnamefont {G.}~\bibnamefont {Hagen}}, \bibinfo
  {author} {\bibfnamefont {G.~R.}\ \bibnamefont {Jansen}}, \bibinfo {author}
  {\bibfnamefont {P.}~\bibnamefont {Navratil}}, \bibinfo {author}
  {\bibfnamefont {F.}~\bibnamefont {Ameil}}, \bibinfo {author} {\bibfnamefont
  {J.}~\bibnamefont {Atkinson}}, \bibinfo {author} {\bibfnamefont
  {Y.}~\bibnamefont {Ayyad}}, \bibinfo {author} {\bibfnamefont
  {D.}~\bibnamefont {Cortina-Gil}}, \bibinfo {author} {\bibfnamefont
  {I.}~\bibnamefont {Dillmann}}, \bibinfo {author} {\bibfnamefont
  {A.}~\bibnamefont {Estrad\'e}}, \bibinfo {author} {\bibfnamefont
  {A.}~\bibnamefont {Evdokimov}}, \bibinfo {author} {\bibfnamefont
  {F.}~\bibnamefont {Farinon}}, \bibinfo {author} {\bibfnamefont
  {H.}~\bibnamefont {Geissel}}, \bibinfo {author} {\bibfnamefont
  {G.}~\bibnamefont {Guastalla}}, \bibinfo {author} {\bibfnamefont
  {R.}~\bibnamefont {Janik}}, \bibinfo {author} {\bibfnamefont
  {M.}~\bibnamefont {Kimura}}, \bibinfo {author} {\bibfnamefont
  {R.}~\bibnamefont {Kn\"obel}}, \bibinfo {author} {\bibfnamefont
  {J.}~\bibnamefont {Kurcewicz}}, \bibinfo {author} {\bibfnamefont {Y.~A.}\
  \bibnamefont {Litvinov}}, \bibinfo {author} {\bibfnamefont {M.}~\bibnamefont
  {Marta}}, \bibinfo {author} {\bibfnamefont {M.}~\bibnamefont {Mostazo}},
  \bibinfo {author} {\bibfnamefont {I.}~\bibnamefont {Mukha}}, \bibinfo
  {author} {\bibfnamefont {C.}~\bibnamefont {Nociforo}}, \bibinfo {author}
  {\bibfnamefont {H.~J.}\ \bibnamefont {Ong}}, \bibinfo {author} {\bibfnamefont
  {S.}~\bibnamefont {Pietri}}, \bibinfo {author} {\bibfnamefont
  {A.}~\bibnamefont {Prochazka}}, \bibinfo {author} {\bibfnamefont
  {C.}~\bibnamefont {Scheidenberger}}, \bibinfo {author} {\bibfnamefont
  {B.}~\bibnamefont {Sitar}}, \bibinfo {author} {\bibfnamefont
  {P.}~\bibnamefont {Strmen}}, \bibinfo {author} {\bibfnamefont
  {Y.}~\bibnamefont {Suzuki}}, \bibinfo {author} {\bibfnamefont
  {M.}~\bibnamefont {Takechi}}, \bibinfo {author} {\bibfnamefont
  {J.}~\bibnamefont {Tanaka}}, \bibinfo {author} {\bibfnamefont
  {I.}~\bibnamefont {Tanihata}}, \bibinfo {author} {\bibfnamefont
  {S.}~\bibnamefont {Terashima}}, \bibinfo {author} {\bibfnamefont
  {J.}~\bibnamefont {Vargas}}, \bibinfo {author} {\bibfnamefont
  {H.}~\bibnamefont {Weick}}, \ and\ \bibinfo {author} {\bibfnamefont {J.~S.}\
  \bibnamefont {Winfield}},\ }\href {\doibase 10.1103/PhysRevLett.117.102501}
  {\bibfield  {journal} {\bibinfo  {journal} {Phys. Rev. Lett.}\ }\textbf
  {\bibinfo {volume} {117}},\ \bibinfo {pages} {102501} (\bibinfo {year}
  {2016})}\BibitemShut {NoStop}%
\bibitem [{\citenamefont {Hu}\ \emph {et~al.}(2019)\citenamefont {Hu},
  \citenamefont {Wu}, \citenamefont {Sun},\ and\ \citenamefont
  {Xu}}]{B.S.Hu-PhysRevC.99.061302(2019)}%
  \BibitemOpen
  \bibfield  {author} {\bibinfo {author} {\bibfnamefont {B.~S.}\ \bibnamefont
  {Hu}}, \bibinfo {author} {\bibfnamefont {Q.}~\bibnamefont {Wu}}, \bibinfo
  {author} {\bibfnamefont {Z.~H.}\ \bibnamefont {Sun}}, \ and\ \bibinfo
  {author} {\bibfnamefont {F.~R.}\ \bibnamefont {Xu}},\ }\href {\doibase
  10.1103/PhysRevC.99.061302} {\bibfield  {journal} {\bibinfo  {journal} {Phys.
  Rev. C}\ }\textbf {\bibinfo {volume} {99}},\ \bibinfo {pages} {061302}
  (\bibinfo {year} {2019})}\BibitemShut {NoStop}%
\bibitem [{\citenamefont {White}(2002)}]{White-J.Chem.Phys.117.7472(2002)}%
  \BibitemOpen
  \bibfield  {author} {\bibinfo {author} {\bibfnamefont {S.~R.}\ \bibnamefont
  {White}},\ }\href {\doibase 10.1063/1.1508370} {\bibfield  {journal}
  {\bibinfo  {journal} {J. Chem. Phys.}\ }\textbf {\bibinfo {volume} {117}},\
  \bibinfo {pages} {7472} (\bibinfo {year} {2002})}\BibitemShut {NoStop}%
\bibitem [{\citenamefont {Morris}\ \emph {et~al.}(2015)\citenamefont {Morris},
  \citenamefont {Parzuchowski},\ and\ \citenamefont
  {Bogner}}]{PhysRevC.92.034331}%
  \BibitemOpen
  \bibfield  {author} {\bibinfo {author} {\bibfnamefont {T.~D.}\ \bibnamefont
  {Morris}}, \bibinfo {author} {\bibfnamefont {N.~M.}\ \bibnamefont
  {Parzuchowski}}, \ and\ \bibinfo {author} {\bibfnamefont {S.~K.}\
  \bibnamefont {Bogner}},\ }\href {\doibase 10.1103/PhysRevC.92.034331}
  {\bibfield  {journal} {\bibinfo  {journal} {Phys. Rev. C}\ }\textbf {\bibinfo
  {volume} {92}},\ \bibinfo {pages} {034331} (\bibinfo {year}
  {2015})}\BibitemShut {NoStop}%
\bibitem [{\citenamefont {Wang}\ \emph {et~al.}(2021)\citenamefont {Wang},
  \citenamefont {Huang}, \citenamefont {Kondev}, \citenamefont {Audi},\ and\
  \citenamefont {Naimi}}]{Wang_2021}%
  \BibitemOpen
  \bibfield  {author} {\bibinfo {author} {\bibfnamefont {M.}~\bibnamefont
  {Wang}}, \bibinfo {author} {\bibfnamefont {W.}~\bibnamefont {Huang}},
  \bibinfo {author} {\bibfnamefont {F.}~\bibnamefont {Kondev}}, \bibinfo
  {author} {\bibfnamefont {G.}~\bibnamefont {Audi}}, \ and\ \bibinfo {author}
  {\bibfnamefont {S.}~\bibnamefont {Naimi}},\ }\href {\doibase
  10.1088/1674-1137/abddaf} {\bibfield  {journal} {\bibinfo  {journal} {Chin.
  Phys. C}\ }\textbf {\bibinfo {volume} {45}},\ \bibinfo {pages} {030003}
  (\bibinfo {year} {2021})}\BibitemShut {NoStop}%
\bibitem [{\citenamefont {Roth}\ and\ \citenamefont
  {Navr\'atil}(2007)}]{R.Roth-PhysRevLett.99.092501(2007)}%
  \BibitemOpen
  \bibfield  {author} {\bibinfo {author} {\bibfnamefont {R.}~\bibnamefont
  {Roth}}\ and\ \bibinfo {author} {\bibfnamefont {P.}~\bibnamefont
  {Navr\'atil}},\ }\href {\doibase 10.1103/PhysRevLett.99.092501} {\bibfield
  {journal} {\bibinfo  {journal} {Phys. Rev. Lett.}\ }\textbf {\bibinfo
  {volume} {99}},\ \bibinfo {pages} {092501} (\bibinfo {year}
  {2007})}\BibitemShut {NoStop}%
\bibitem [{\citenamefont {Roth}(2009)}]{R.Roth-PhysRevC.79.064324(2009)}%
  \BibitemOpen
  \bibfield  {author} {\bibinfo {author} {\bibfnamefont {R.}~\bibnamefont
  {Roth}},\ }\href {\doibase 10.1103/PhysRevC.79.064324} {\bibfield  {journal}
  {\bibinfo  {journal} {Phys. Rev. C}\ }\textbf {\bibinfo {volume} {79}},\
  \bibinfo {pages} {064324} (\bibinfo {year} {2009})}\BibitemShut {NoStop}%
\bibitem [{\citenamefont {Roth}\ \emph {et~al.}(2011)\citenamefont {Roth},
  \citenamefont {Langhammer}, \citenamefont {Calci}, \citenamefont {Binder},\
  and\ \citenamefont {Navr\'atil}}]{R.Roth-PhysRevLett.107.072501(2011)}%
  \BibitemOpen
  \bibfield  {author} {\bibinfo {author} {\bibfnamefont {R.}~\bibnamefont
  {Roth}}, \bibinfo {author} {\bibfnamefont {J.}~\bibnamefont {Langhammer}},
  \bibinfo {author} {\bibfnamefont {A.}~\bibnamefont {Calci}}, \bibinfo
  {author} {\bibfnamefont {S.}~\bibnamefont {Binder}}, \ and\ \bibinfo {author}
  {\bibfnamefont {P.}~\bibnamefont {Navr\'atil}},\ }\href {\doibase
  10.1103/PhysRevLett.107.072501} {\bibfield  {journal} {\bibinfo  {journal}
  {Phys. Rev. Lett.}\ }\textbf {\bibinfo {volume} {107}},\ \bibinfo {pages}
  {072501} (\bibinfo {year} {2011})}\BibitemShut {NoStop}%
\bibitem [{\citenamefont {Caprio}\ \emph {et~al.}(2015)\citenamefont {Caprio},
  \citenamefont {Maris}, \citenamefont {Vary},\ and\ \citenamefont
  {Smith}}]{M.A.Caprio-J.Mod.Phys.E.24.1541002(2015)}%
  \BibitemOpen
  \bibfield  {author} {\bibinfo {author} {\bibfnamefont {M.~A.}\ \bibnamefont
  {Caprio}}, \bibinfo {author} {\bibfnamefont {P.}~\bibnamefont {Maris}},
  \bibinfo {author} {\bibfnamefont {J.~P.}\ \bibnamefont {Vary}}, \ and\
  \bibinfo {author} {\bibfnamefont {R.}~\bibnamefont {Smith}},\ }\href
  {\doibase 10.1142/s0218301315410025} {\bibfield  {journal} {\bibinfo
  {journal} {Int. J. Mod. Phys. E}\ }\textbf {\bibinfo {volume} {24}},\
  \bibinfo {pages} {1541002} (\bibinfo {year} {2015})}\BibitemShut {NoStop}%
\bibitem [{\citenamefont {Abe}\ \emph {et~al.}(2021)\citenamefont {Abe},
  \citenamefont {Maris}, \citenamefont {Otsuka}, \citenamefont {Shimizu},
  \citenamefont {Utsuno},\ and\ \citenamefont
  {Vary}}]{T.Abe-PhysRevC.104.054315(2021)}%
  \BibitemOpen
  \bibfield  {author} {\bibinfo {author} {\bibfnamefont {T.}~\bibnamefont
  {Abe}}, \bibinfo {author} {\bibfnamefont {P.}~\bibnamefont {Maris}}, \bibinfo
  {author} {\bibfnamefont {T.}~\bibnamefont {Otsuka}}, \bibinfo {author}
  {\bibfnamefont {N.}~\bibnamefont {Shimizu}}, \bibinfo {author} {\bibfnamefont
  {Y.}~\bibnamefont {Utsuno}}, \ and\ \bibinfo {author} {\bibfnamefont {J.~P.}\
  \bibnamefont {Vary}},\ }\href {\doibase 10.1103/PhysRevC.104.054315}
  {\bibfield  {journal} {\bibinfo  {journal} {Phys. Rev. C}\ }\textbf {\bibinfo
  {volume} {104}},\ \bibinfo {pages} {054315} (\bibinfo {year}
  {2021})}\BibitemShut {NoStop}%
\bibitem [{\citenamefont {Henderson}\ \emph {et~al.}(2018)\citenamefont
  {Henderson}, \citenamefont {Hackman}, \citenamefont {Ruotsalainen},
  \citenamefont {Stroberg}, \citenamefont {Launey}, \citenamefont {Holt},
  \citenamefont {Ali}, \citenamefont {Bernier}, \citenamefont {Bentley},
  \citenamefont {Bowry}, \citenamefont {Caballero-Folch}, \citenamefont
  {Evitts}, \citenamefont {Frederick}, \citenamefont {Garnsworthy},
  \citenamefont {Garrett}, \citenamefont {Jigmeddorj}, \citenamefont {Kilic},
  \citenamefont {Lassen}, \citenamefont {Measures}, \citenamefont {Muecher},
  \citenamefont {Olaizola}, \citenamefont {O'Sullivan}, \citenamefont
  {Paetkau}, \citenamefont {Park}, \citenamefont {Smallcombe}, \citenamefont
  {Svensson}, \citenamefont {Wadsworth},\ and\ \citenamefont
  {Wu}}]{J.Henderson-Phys.Lett.B.782.468(2018)}%
  \BibitemOpen
  \bibfield  {author} {\bibinfo {author} {\bibfnamefont {J.}~\bibnamefont
  {Henderson}}, \bibinfo {author} {\bibfnamefont {G.}~\bibnamefont {Hackman}},
  \bibinfo {author} {\bibfnamefont {P.}~\bibnamefont {Ruotsalainen}}, \bibinfo
  {author} {\bibfnamefont {S.}~\bibnamefont {Stroberg}}, \bibinfo {author}
  {\bibfnamefont {K.}~\bibnamefont {Launey}}, \bibinfo {author} {\bibfnamefont
  {J.}~\bibnamefont {Holt}}, \bibinfo {author} {\bibfnamefont {F.}~\bibnamefont
  {Ali}}, \bibinfo {author} {\bibfnamefont {N.}~\bibnamefont {Bernier}},
  \bibinfo {author} {\bibfnamefont {M.}~\bibnamefont {Bentley}}, \bibinfo
  {author} {\bibfnamefont {M.}~\bibnamefont {Bowry}}, \bibinfo {author}
  {\bibfnamefont {R.}~\bibnamefont {Caballero-Folch}}, \bibinfo {author}
  {\bibfnamefont {L.}~\bibnamefont {Evitts}}, \bibinfo {author} {\bibfnamefont
  {R.}~\bibnamefont {Frederick}}, \bibinfo {author} {\bibfnamefont
  {A.}~\bibnamefont {Garnsworthy}}, \bibinfo {author} {\bibfnamefont
  {P.}~\bibnamefont {Garrett}}, \bibinfo {author} {\bibfnamefont
  {B.}~\bibnamefont {Jigmeddorj}}, \bibinfo {author} {\bibfnamefont
  {A.}~\bibnamefont {Kilic}}, \bibinfo {author} {\bibfnamefont
  {J.}~\bibnamefont {Lassen}}, \bibinfo {author} {\bibfnamefont
  {J.}~\bibnamefont {Measures}}, \bibinfo {author} {\bibfnamefont
  {D.}~\bibnamefont {Muecher}}, \bibinfo {author} {\bibfnamefont
  {B.}~\bibnamefont {Olaizola}}, \bibinfo {author} {\bibfnamefont
  {E.}~\bibnamefont {O'Sullivan}}, \bibinfo {author} {\bibfnamefont
  {O.}~\bibnamefont {Paetkau}}, \bibinfo {author} {\bibfnamefont
  {J.}~\bibnamefont {Park}}, \bibinfo {author} {\bibfnamefont {J.}~\bibnamefont
  {Smallcombe}}, \bibinfo {author} {\bibfnamefont {C.}~\bibnamefont
  {Svensson}}, \bibinfo {author} {\bibfnamefont {R.}~\bibnamefont {Wadsworth}},
  \ and\ \bibinfo {author} {\bibfnamefont {C.}~\bibnamefont {Wu}},\ }\href
  {\doibase https://doi.org/10.1016/j.physletb.2018.05.064} {\bibfield
  {journal} {\bibinfo  {journal} {Phys. Lett. B}\ }\textbf {\bibinfo {volume}
  {782}},\ \bibinfo {pages} {468} (\bibinfo {year} {2018})}\BibitemShut
  {NoStop}%
\bibitem [{\citenamefont {Henderson}\ \emph {et~al.}(2020)\citenamefont
  {Henderson}, \citenamefont {Hackman}, \citenamefont {Ruotsalainen},
  \citenamefont {Holt}, \citenamefont {Stroberg}, \citenamefont {Hagen},
  \citenamefont {Andreoiu}, \citenamefont {Ball}, \citenamefont {Bernier},
  \citenamefont {Bowry}, \citenamefont {Caballero-Folch}, \citenamefont {Cruz},
  \citenamefont {Varela}, \citenamefont {Evitts}, \citenamefont {Frederick},
  \citenamefont {Garnsworthy}, \citenamefont {Holl}, \citenamefont {Lassen},
  \citenamefont {Measures}, \citenamefont {Olaizola}, \citenamefont
  {O'Sullivan}, \citenamefont {Paetkau}, \citenamefont {Park}, \citenamefont
  {Smallcombe}, \citenamefont {Svensson}, \citenamefont {Whitmore},\ and\
  \citenamefont {Wu}}]{J.Henderson-arxiv.2005.03796(2020)}%
  \BibitemOpen
  \bibfield  {author} {\bibinfo {author} {\bibfnamefont {J.}~\bibnamefont
  {Henderson}}, \bibinfo {author} {\bibfnamefont {G.}~\bibnamefont {Hackman}},
  \bibinfo {author} {\bibfnamefont {P.}~\bibnamefont {Ruotsalainen}}, \bibinfo
  {author} {\bibfnamefont {J.~D.}\ \bibnamefont {Holt}}, \bibinfo {author}
  {\bibfnamefont {S.~R.}\ \bibnamefont {Stroberg}}, \bibinfo {author}
  {\bibfnamefont {G.}~\bibnamefont {Hagen}}, \bibinfo {author} {\bibfnamefont
  {C.}~\bibnamefont {Andreoiu}}, \bibinfo {author} {\bibfnamefont {G.~C.}\
  \bibnamefont {Ball}}, \bibinfo {author} {\bibfnamefont {N.}~\bibnamefont
  {Bernier}}, \bibinfo {author} {\bibfnamefont {M.}~\bibnamefont {Bowry}},
  \bibinfo {author} {\bibfnamefont {R.}~\bibnamefont {Caballero-Folch}},
  \bibinfo {author} {\bibfnamefont {S.}~\bibnamefont {Cruz}}, \bibinfo {author}
  {\bibfnamefont {A.~D.}\ \bibnamefont {Varela}}, \bibinfo {author}
  {\bibfnamefont {L.~J.}\ \bibnamefont {Evitts}}, \bibinfo {author}
  {\bibfnamefont {R.}~\bibnamefont {Frederick}}, \bibinfo {author}
  {\bibfnamefont {A.~B.}\ \bibnamefont {Garnsworthy}}, \bibinfo {author}
  {\bibfnamefont {M.}~\bibnamefont {Holl}}, \bibinfo {author} {\bibfnamefont
  {J.}~\bibnamefont {Lassen}}, \bibinfo {author} {\bibfnamefont
  {J.}~\bibnamefont {Measures}}, \bibinfo {author} {\bibfnamefont
  {B.}~\bibnamefont {Olaizola}}, \bibinfo {author} {\bibfnamefont
  {E.}~\bibnamefont {O'Sullivan}}, \bibinfo {author} {\bibfnamefont
  {O.}~\bibnamefont {Paetkau}}, \bibinfo {author} {\bibfnamefont
  {J.}~\bibnamefont {Park}}, \bibinfo {author} {\bibfnamefont {J.}~\bibnamefont
  {Smallcombe}}, \bibinfo {author} {\bibfnamefont {C.~E.}\ \bibnamefont
  {Svensson}}, \bibinfo {author} {\bibfnamefont {K.}~\bibnamefont {Whitmore}},
  \ and\ \bibinfo {author} {\bibfnamefont {C.~Y.}\ \bibnamefont {Wu}},\
  }\href@noop {} {} (\bibinfo {year} {2020}),\ \Eprint
  {http://arxiv.org/abs/2005.03796} {arXiv:2005.03796 [nucl-ex]} \BibitemShut
  {NoStop}%
\bibitem [{\citenamefont {Krieger}\ \emph {et~al.}(2016)\citenamefont
  {Krieger}, \citenamefont {Nörtershäuser}, \citenamefont {Geppert},
  \citenamefont {Blaum}, \citenamefont {Bissell}, \citenamefont {Frömmgen},
  \citenamefont {Hammen}, \citenamefont {Kreim}, \citenamefont {Kowalska},
  \citenamefont {Krämer}, \citenamefont {Neugart}, \citenamefont {Neyens},
  \citenamefont {Sánchez}, \citenamefont {Tiedemann}, \citenamefont
  {Yordanov},\ and\ \citenamefont
  {Zakova}}]{A.Krieger-Appl.Phys.B.123.0(2016)}%
  \BibitemOpen
  \bibfield  {author} {\bibinfo {author} {\bibfnamefont {A.}~\bibnamefont
  {Krieger}}, \bibinfo {author} {\bibfnamefont {W.}~\bibnamefont
  {Nörtershäuser}}, \bibinfo {author} {\bibfnamefont {C.}~\bibnamefont
  {Geppert}}, \bibinfo {author} {\bibfnamefont {K.}~\bibnamefont {Blaum}},
  \bibinfo {author} {\bibfnamefont {M.~L.}\ \bibnamefont {Bissell}}, \bibinfo
  {author} {\bibfnamefont {N.}~\bibnamefont {Frömmgen}}, \bibinfo {author}
  {\bibfnamefont {M.}~\bibnamefont {Hammen}}, \bibinfo {author} {\bibfnamefont
  {K.}~\bibnamefont {Kreim}}, \bibinfo {author} {\bibfnamefont
  {M.}~\bibnamefont {Kowalska}}, \bibinfo {author} {\bibfnamefont
  {J.}~\bibnamefont {Krämer}}, \bibinfo {author} {\bibfnamefont
  {R.}~\bibnamefont {Neugart}}, \bibinfo {author} {\bibfnamefont
  {G.}~\bibnamefont {Neyens}}, \bibinfo {author} {\bibfnamefont
  {R.}~\bibnamefont {Sánchez}}, \bibinfo {author} {\bibfnamefont
  {D.}~\bibnamefont {Tiedemann}}, \bibinfo {author} {\bibfnamefont {D.~T.}\
  \bibnamefont {Yordanov}}, \ and\ \bibinfo {author} {\bibfnamefont
  {M.}~\bibnamefont {Zakova}},\ }\href {\doibase 10.1007/s00340-016-6579-5}
  {\bibfield  {journal} {\bibinfo  {journal} {Appl. Phys. B}\ }\textbf
  {\bibinfo {volume} {123}},\ \bibinfo {pages} {0} (\bibinfo {year}
  {2016})}\BibitemShut {NoStop}%
\bibitem [{\citenamefont {Lapoux}\ \emph {et~al.}(2016)\citenamefont {Lapoux},
  \citenamefont {Som\`a}, \citenamefont {Barbieri}, \citenamefont {Hergert},
  \citenamefont {Holt},\ and\ \citenamefont
  {Stroberg}}]{V.Lapoux-PhysRevLett.117.052501(2016)}%
  \BibitemOpen
  \bibfield  {author} {\bibinfo {author} {\bibfnamefont {V.}~\bibnamefont
  {Lapoux}}, \bibinfo {author} {\bibfnamefont {V.}~\bibnamefont {Som\`a}},
  \bibinfo {author} {\bibfnamefont {C.}~\bibnamefont {Barbieri}}, \bibinfo
  {author} {\bibfnamefont {H.}~\bibnamefont {Hergert}}, \bibinfo {author}
  {\bibfnamefont {J.~D.}\ \bibnamefont {Holt}}, \ and\ \bibinfo {author}
  {\bibfnamefont {S.~R.}\ \bibnamefont {Stroberg}},\ }\href {\doibase
  10.1103/PhysRevLett.117.052501} {\bibfield  {journal} {\bibinfo  {journal}
  {Phys. Rev. Lett.}\ }\textbf {\bibinfo {volume} {117}},\ \bibinfo {pages}
  {052501} (\bibinfo {year} {2016})}\BibitemShut {NoStop}%
\bibitem [{\citenamefont {Ohayon}\ \emph {et~al.}(2019)\citenamefont {Ohayon},
  \citenamefont {Rahangdale}, \citenamefont {Geddes}, \citenamefont
  {Berengut},\ and\ \citenamefont {Ron}}]{B.Ohayon-PhysRevA.99.042503(2019)}%
  \BibitemOpen
  \bibfield  {author} {\bibinfo {author} {\bibfnamefont {B.}~\bibnamefont
  {Ohayon}}, \bibinfo {author} {\bibfnamefont {H.}~\bibnamefont {Rahangdale}},
  \bibinfo {author} {\bibfnamefont {A.~J.}\ \bibnamefont {Geddes}}, \bibinfo
  {author} {\bibfnamefont {J.~C.}\ \bibnamefont {Berengut}}, \ and\ \bibinfo
  {author} {\bibfnamefont {G.}~\bibnamefont {Ron}},\ }\href {\doibase
  10.1103/PhysRevA.99.042503} {\bibfield  {journal} {\bibinfo  {journal} {Phys.
  Rev. A}\ }\textbf {\bibinfo {volume} {99}},\ \bibinfo {pages} {042503}
  (\bibinfo {year} {2019})}\BibitemShut {NoStop}%
\bibitem [{\citenamefont {Yordanov}\ \emph {et~al.}(2012)\citenamefont
  {Yordanov}, \citenamefont {Bissell}, \citenamefont {Blaum}, \citenamefont
  {De~Rydt}, \citenamefont {Geppert}, \citenamefont {Kowalska}, \citenamefont
  {Kr\"amer}, \citenamefont {Kreim}, \citenamefont {Krieger}, \citenamefont
  {Lievens}, \citenamefont {Neff}, \citenamefont {Neugart}, \citenamefont
  {Neyens}, \citenamefont {N\"ortersh\"auser}, \citenamefont {S\'anchez},\ and\
  \citenamefont {Vingerhoets}}]{D.Yordanov-PhysRevLett.108.042504(2012)}%
  \BibitemOpen
  \bibfield  {author} {\bibinfo {author} {\bibfnamefont {D.~T.}\ \bibnamefont
  {Yordanov}}, \bibinfo {author} {\bibfnamefont {M.~L.}\ \bibnamefont
  {Bissell}}, \bibinfo {author} {\bibfnamefont {K.}~\bibnamefont {Blaum}},
  \bibinfo {author} {\bibfnamefont {M.}~\bibnamefont {De~Rydt}}, \bibinfo
  {author} {\bibfnamefont {C.}~\bibnamefont {Geppert}}, \bibinfo {author}
  {\bibfnamefont {M.}~\bibnamefont {Kowalska}}, \bibinfo {author}
  {\bibfnamefont {J.}~\bibnamefont {Kr\"amer}}, \bibinfo {author}
  {\bibfnamefont {K.}~\bibnamefont {Kreim}}, \bibinfo {author} {\bibfnamefont
  {A.}~\bibnamefont {Krieger}}, \bibinfo {author} {\bibfnamefont
  {P.}~\bibnamefont {Lievens}}, \bibinfo {author} {\bibfnamefont
  {T.}~\bibnamefont {Neff}}, \bibinfo {author} {\bibfnamefont {R.}~\bibnamefont
  {Neugart}}, \bibinfo {author} {\bibfnamefont {G.}~\bibnamefont {Neyens}},
  \bibinfo {author} {\bibfnamefont {W.}~\bibnamefont {N\"ortersh\"auser}},
  \bibinfo {author} {\bibfnamefont {R.}~\bibnamefont {S\'anchez}}, \ and\
  \bibinfo {author} {\bibfnamefont {P.}~\bibnamefont {Vingerhoets}},\ }\href
  {\doibase 10.1103/PhysRevLett.108.042504} {\bibfield  {journal} {\bibinfo
  {journal} {Phys. Rev. Lett.}\ }\textbf {\bibinfo {volume} {108}},\ \bibinfo
  {pages} {042504} (\bibinfo {year} {2012})}\BibitemShut {NoStop}%
\bibitem [{\citenamefont {Stroberg}\ \emph {et~al.}(2021)\citenamefont
  {Stroberg}, \citenamefont {Holt}, \citenamefont {Schwenk},\ and\
  \citenamefont {Simonis}}]{S.R.Stroberg-Phys.Rev.Lett.126.022501(2021)}%
  \BibitemOpen
  \bibfield  {author} {\bibinfo {author} {\bibfnamefont {S.~R.}\ \bibnamefont
  {Stroberg}}, \bibinfo {author} {\bibfnamefont {J.~D.}\ \bibnamefont {Holt}},
  \bibinfo {author} {\bibfnamefont {A.}~\bibnamefont {Schwenk}}, \ and\
  \bibinfo {author} {\bibfnamefont {J.}~\bibnamefont {Simonis}},\ }\href
  {\doibase 10.1103/PhysRevLett.126.022501} {\bibfield  {journal} {\bibinfo
  {journal} {Phys. Rev. Lett.}\ }\textbf {\bibinfo {volume} {126}},\ \bibinfo
  {pages} {022501} (\bibinfo {year} {2021})}\BibitemShut {NoStop}%
\bibitem [{\citenamefont {Ma}\ \emph {et~al.}(2020{\natexlab{a}})\citenamefont
  {Ma}, \citenamefont {Xu}, \citenamefont {Coraggio}, \citenamefont {Hu},
  \citenamefont {Li}, \citenamefont {Fukui}, \citenamefont {{De Angelis}},
  \citenamefont {Itaco},\ and\ \citenamefont
  {Gargano}}]{Y.Z.Ma-Phys.Lett.B.802.135257(2020)}%
  \BibitemOpen
  \bibfield  {author} {\bibinfo {author} {\bibfnamefont {Y.}~\bibnamefont
  {Ma}}, \bibinfo {author} {\bibfnamefont {F.}~\bibnamefont {Xu}}, \bibinfo
  {author} {\bibfnamefont {L.}~\bibnamefont {Coraggio}}, \bibinfo {author}
  {\bibfnamefont {B.}~\bibnamefont {Hu}}, \bibinfo {author} {\bibfnamefont
  {J.}~\bibnamefont {Li}}, \bibinfo {author} {\bibfnamefont {T.}~\bibnamefont
  {Fukui}}, \bibinfo {author} {\bibfnamefont {L.}~\bibnamefont {{De Angelis}}},
  \bibinfo {author} {\bibfnamefont {N.}~\bibnamefont {Itaco}}, \ and\ \bibinfo
  {author} {\bibfnamefont {A.}~\bibnamefont {Gargano}},\ }\href {\doibase
  https://doi.org/10.1016/j.physletb.2020.135257} {\bibfield  {journal}
  {\bibinfo  {journal} {Phys. Lett. B}\ }\textbf {\bibinfo {volume} {802}},\
  \bibinfo {pages} {135257} (\bibinfo {year} {2020}{\natexlab{a}})}\BibitemShut
  {NoStop}%
\bibitem [{\citenamefont {Ma}\ \emph {et~al.}(2020{\natexlab{b}})\citenamefont
  {Ma}, \citenamefont {Xu}, \citenamefont {Michel}, \citenamefont {Zhang},
  \citenamefont {Li}, \citenamefont {Hu}, \citenamefont {Coraggio},
  \citenamefont {Itaco},\ and\ \citenamefont
  {Gargano}}]{Y.Z.Ma-Phys.Lett.B.808.135673(2020)}%
  \BibitemOpen
  \bibfield  {author} {\bibinfo {author} {\bibfnamefont {Y.~Z.}\ \bibnamefont
  {Ma}}, \bibinfo {author} {\bibfnamefont {F.~R.}\ \bibnamefont {Xu}}, \bibinfo
  {author} {\bibfnamefont {N.}~\bibnamefont {Michel}}, \bibinfo {author}
  {\bibfnamefont {S.}~\bibnamefont {Zhang}}, \bibinfo {author} {\bibfnamefont
  {J.~G.}\ \bibnamefont {Li}}, \bibinfo {author} {\bibfnamefont {B.~S.}\
  \bibnamefont {Hu}}, \bibinfo {author} {\bibfnamefont {L.}~\bibnamefont
  {Coraggio}}, \bibinfo {author} {\bibfnamefont {N.}~\bibnamefont {Itaco}}, \
  and\ \bibinfo {author} {\bibfnamefont {A.}~\bibnamefont {Gargano}},\ }\href
  {\doibase https://doi.org/10.1016/j.physletb.2020.135673} {\bibfield
  {journal} {\bibinfo  {journal} {Phys. Lett. B}\ }\textbf {\bibinfo {volume}
  {808}},\ \bibinfo {pages} {135673} (\bibinfo {year}
  {2020}{\natexlab{b}})}\BibitemShut {NoStop}%
\bibitem [{\citenamefont {Li}\ \emph {et~al.}(2020)\citenamefont {Li},
  \citenamefont {Hu}, \citenamefont {Wu}, \citenamefont {Gao}, \citenamefont
  {Dai},\ and\ \citenamefont {Xu}}]{J.G.Li-PhysRevC.102.034302(2020)}%
  \BibitemOpen
  \bibfield  {author} {\bibinfo {author} {\bibfnamefont {J.~G.}\ \bibnamefont
  {Li}}, \bibinfo {author} {\bibfnamefont {B.~S.}\ \bibnamefont {Hu}}, \bibinfo
  {author} {\bibfnamefont {Q.}~\bibnamefont {Wu}}, \bibinfo {author}
  {\bibfnamefont {Y.}~\bibnamefont {Gao}}, \bibinfo {author} {\bibfnamefont
  {S.~J.}\ \bibnamefont {Dai}}, \ and\ \bibinfo {author} {\bibfnamefont
  {F.~R.}\ \bibnamefont {Xu}},\ }\href {\doibase 10.1103/PhysRevC.102.034302}
  {\bibfield  {journal} {\bibinfo  {journal} {Phys. Rev. C}\ }\textbf {\bibinfo
  {volume} {102}},\ \bibinfo {pages} {034302} (\bibinfo {year}
  {2020})}\BibitemShut {NoStop}%
\bibitem [{\citenamefont {Hagen}\ \emph
  {et~al.}(2016{\natexlab{a}})\citenamefont {Hagen}, \citenamefont
  {Hjorth-Jensen}, \citenamefont {Jansen},\ and\ \citenamefont
  {Papenbrock}}]{G.Hagen-Phys.Scr.91.063006(2016)}%
  \BibitemOpen
  \bibfield  {author} {\bibinfo {author} {\bibfnamefont {G.}~\bibnamefont
  {Hagen}}, \bibinfo {author} {\bibfnamefont {M.}~\bibnamefont
  {Hjorth-Jensen}}, \bibinfo {author} {\bibfnamefont {G.~R.}\ \bibnamefont
  {Jansen}}, \ and\ \bibinfo {author} {\bibfnamefont {T.}~\bibnamefont
  {Papenbrock}},\ }\href {\doibase 10.1088/0031-8949/91/6/063006} {\bibfield
  {journal} {\bibinfo  {journal} {Phys. Scr.}\ }\textbf {\bibinfo {volume}
  {91}},\ \bibinfo {pages} {063006} (\bibinfo {year}
  {2016}{\natexlab{a}})}\BibitemShut {NoStop}%
\bibitem [{\citenamefont {Poves}\ and\ \citenamefont
  {Retamosa}(1987)}]{A.Poves-Phys.Lett.B.184.311(1987)}%
  \BibitemOpen
  \bibfield  {author} {\bibinfo {author} {\bibfnamefont {A.}~\bibnamefont
  {Poves}}\ and\ \bibinfo {author} {\bibfnamefont {J.}~\bibnamefont
  {Retamosa}},\ }\href {\doibase https://doi.org/10.1016/0370-2693(87)90171-7}
  {\bibfield  {journal} {\bibinfo  {journal} {Phys. Lett. B}\ }\textbf
  {\bibinfo {volume} {184}},\ \bibinfo {pages} {311} (\bibinfo {year}
  {1987})}\BibitemShut {NoStop}%
\bibitem [{\citenamefont {Warburton}\ \emph {et~al.}(1990)\citenamefont
  {Warburton}, \citenamefont {Becker},\ and\ \citenamefont
  {Brown}}]{E.K.Warburton-PhysRevC.41.1147(1990)}%
  \BibitemOpen
  \bibfield  {author} {\bibinfo {author} {\bibfnamefont {E.~K.}\ \bibnamefont
  {Warburton}}, \bibinfo {author} {\bibfnamefont {J.~A.}\ \bibnamefont
  {Becker}}, \ and\ \bibinfo {author} {\bibfnamefont {B.~A.}\ \bibnamefont
  {Brown}},\ }\href {\doibase 10.1103/PhysRevC.41.1147} {\bibfield  {journal}
  {\bibinfo  {journal} {Phys. Rev. C}\ }\textbf {\bibinfo {volume} {41}},\
  \bibinfo {pages} {1147} (\bibinfo {year} {1990})}\BibitemShut {NoStop}%
\bibitem [{\citenamefont {Miyagi}\ \emph {et~al.}(2020)\citenamefont {Miyagi},
  \citenamefont {Stroberg}, \citenamefont {Holt},\ and\ \citenamefont
  {Shimizu}}]{T.Miyagi-PhysRevC.102.034320(2020)}%
  \BibitemOpen
  \bibfield  {author} {\bibinfo {author} {\bibfnamefont {T.}~\bibnamefont
  {Miyagi}}, \bibinfo {author} {\bibfnamefont {S.~R.}\ \bibnamefont
  {Stroberg}}, \bibinfo {author} {\bibfnamefont {J.~D.}\ \bibnamefont {Holt}},
  \ and\ \bibinfo {author} {\bibfnamefont {N.}~\bibnamefont {Shimizu}},\ }\href
  {\doibase 10.1103/PhysRevC.102.034320} {\bibfield  {journal} {\bibinfo
  {journal} {Phys. Rev. C}\ }\textbf {\bibinfo {volume} {102}},\ \bibinfo
  {pages} {034320} (\bibinfo {year} {2020})}\BibitemShut {NoStop}%
\bibitem [{\citenamefont {Hagen}\ \emph
  {et~al.}(2016{\natexlab{b}})\citenamefont {Hagen}, \citenamefont {kström},
  \citenamefont {Forssén}, \citenamefont {Jansen}, \citenamefont {Nazarewicz},
  \citenamefont {Papenbrock}, \citenamefont {Wendt}, \citenamefont {Bacca},
  \citenamefont {Barnea}, \citenamefont {Carlsson}, \citenamefont {Drischler},
  \citenamefont {Hebeler}, \citenamefont {Hjorth-Jensen}, \citenamefont
  {Miorelli}, \citenamefont {Orlandini}, \citenamefont {Schwenk},\ and\
  \citenamefont {Simonis}}]{G.Hagen-Nat.Phys.12.186(2016)}%
  \BibitemOpen
  \bibfield  {author} {\bibinfo {author} {\bibfnamefont {G.}~\bibnamefont
  {Hagen}}, \bibinfo {author} {\bibfnamefont {A.}~\bibnamefont {kström}},
  \bibinfo {author} {\bibfnamefont {C.}~\bibnamefont {Forssén}}, \bibinfo
  {author} {\bibfnamefont {G.~R.}\ \bibnamefont {Jansen}}, \bibinfo {author}
  {\bibfnamefont {W.}~\bibnamefont {Nazarewicz}}, \bibinfo {author}
  {\bibfnamefont {T.}~\bibnamefont {Papenbrock}}, \bibinfo {author}
  {\bibfnamefont {K.~A.}\ \bibnamefont {Wendt}}, \bibinfo {author}
  {\bibfnamefont {S.}~\bibnamefont {Bacca}}, \bibinfo {author} {\bibfnamefont
  {N.}~\bibnamefont {Barnea}}, \bibinfo {author} {\bibfnamefont
  {B.}~\bibnamefont {Carlsson}}, \bibinfo {author} {\bibfnamefont
  {C.}~\bibnamefont {Drischler}}, \bibinfo {author} {\bibfnamefont
  {K.}~\bibnamefont {Hebeler}}, \bibinfo {author} {\bibfnamefont
  {M.}~\bibnamefont {Hjorth-Jensen}}, \bibinfo {author} {\bibfnamefont
  {M.}~\bibnamefont {Miorelli}}, \bibinfo {author} {\bibfnamefont
  {G.}~\bibnamefont {Orlandini}}, \bibinfo {author} {\bibfnamefont
  {A.}~\bibnamefont {Schwenk}}, \ and\ \bibinfo {author} {\bibfnamefont
  {J.}~\bibnamefont {Simonis}},\ }\href {\doibase
  https://doi.org/10.1038/nphys3529} {\bibfield  {journal} {\bibinfo  {journal}
  {Nat. Phys.}\ }\textbf {\bibinfo {volume} {12}},\ \bibinfo {pages} {186}
  (\bibinfo {year} {2016}{\natexlab{b}})}\BibitemShut {NoStop}%
\bibitem [{\citenamefont {Bogner}\ \emph {et~al.}(2008)\citenamefont {Bogner},
  \citenamefont {Furnstahl}, \citenamefont {Maris}, \citenamefont {Perry},
  \citenamefont {Schwenk},\ and\ \citenamefont
  {Vary}}]{S.K.Bogner-Nucl.Phys.A.801.21(2008)}%
  \BibitemOpen
  \bibfield  {author} {\bibinfo {author} {\bibfnamefont {S.}~\bibnamefont
  {Bogner}}, \bibinfo {author} {\bibfnamefont {R.}~\bibnamefont {Furnstahl}},
  \bibinfo {author} {\bibfnamefont {P.}~\bibnamefont {Maris}}, \bibinfo
  {author} {\bibfnamefont {R.}~\bibnamefont {Perry}}, \bibinfo {author}
  {\bibfnamefont {A.}~\bibnamefont {Schwenk}}, \ and\ \bibinfo {author}
  {\bibfnamefont {J.}~\bibnamefont {Vary}},\ }\href {\doibase
  https://doi.org/10.1016/j.nuclphysa.2007.12.008} {\bibfield  {journal}
  {\bibinfo  {journal} {Nucl. Phys. A}\ }\textbf {\bibinfo {volume} {801}},\
  \bibinfo {pages} {21} (\bibinfo {year} {2008})}\BibitemShut {NoStop}%
\bibitem [{\citenamefont {Hoppe}\ \emph {et~al.}(2019)\citenamefont {Hoppe},
  \citenamefont {Drischler}, \citenamefont {Hebeler}, \citenamefont {Schwenk},\
  and\ \citenamefont {Simonis}}]{J.Hoppe-PhysRevC.100.024318(2019)}%
  \BibitemOpen
  \bibfield  {author} {\bibinfo {author} {\bibfnamefont {J.}~\bibnamefont
  {Hoppe}}, \bibinfo {author} {\bibfnamefont {C.}~\bibnamefont {Drischler}},
  \bibinfo {author} {\bibfnamefont {K.}~\bibnamefont {Hebeler}}, \bibinfo
  {author} {\bibfnamefont {A.}~\bibnamefont {Schwenk}}, \ and\ \bibinfo
  {author} {\bibfnamefont {J.}~\bibnamefont {Simonis}},\ }\href {\doibase
  10.1103/PhysRevC.100.024318} {\bibfield  {journal} {\bibinfo  {journal}
  {Phys. Rev. C}\ }\textbf {\bibinfo {volume} {100}},\ \bibinfo {pages}
  {024318} (\bibinfo {year} {2019})}\BibitemShut {NoStop}%
\end{thebibliography}%

\end{document}